\documentclass[iop,revtex4]{emulateapj}

\newcommand{\msun} {${\cal M}_\odot$}

\begin{document}

\title{Binary stars in Upper Scorpius}

\author{Andrei Tokovinin}
\affiliation{Cerro Tololo Inter-American Observatory, Casilla 603, La Serena, Chile}
\author{Cesar Brice\~no}
\affiliation{Cerro Tololo Inter-American Observatory, Casilla 603, La Serena, Chile}

\email{atokovinin@ctio.noao.edu}

\begin{abstract}
To  address the  statistics of  binary stars  in the  8-Myr  old Upper
Scorpius star  formation region, we  conducted speckle-interferometric
survey  of  614  association  members  more  massive  than  0.4  \msun
~(spectral types earlier than M3V) based  on the list of Luhman et al.
(2018).   We resolved  187 pairs,  55  of those  are new  discoveries.
Using also the published data and the {\em Gaia} DR2, a catalog of 250
physical  binaries  is  produced.   We carefully  estimated  detection
limits for each target and studied binary statistics in the separation
range  from  0\farcs06  to 20\arcsec  ~(9  to  2800  au), as  well  as
clustering at  larger separations.   The frequency of  companions with
mass ratios $q>0.3$  in this separation range is 0.33$\pm$0.04 and
  0.35$\pm$0.04 for early M and solar-type stars, respectively, larger
  by 1.62$\pm$0.22 and 1.39$\pm$0.18  times compared to field stars of
  similar masses. The  excess is produced mostly by  pairs closer than
  100  au.   At  separations from  100 to  10$^4$ au,  the separation
distribution and companion fraction resemble those of solar-type stars
in the  field.  However, unlike in  the field, we   see a relative
  deficit  of equal-mass binaries   at  separations of  $\sim$500 au,
compared to  smaller and larger separations.  The  distribution of $q$
depends  on the  separation, with  a preference  of larger  $q$  and a
larger fraction  of twins with  $q>0.95$ at smaller  separations.  The
binary population of Upper Scorpius differs from binaries in the field
in  several ways  and suggests   that  binary statistics  is not
universal.
\end{abstract}

\keywords{stars:binary; stars:young}

\section{Introduction}
\label{sec:intro}

Several surveys of  binary stars in young stellar populations have
been conducted since the 1990s.   Their main goal has been to document
differences in binary statistics as  a function of density and age and
to compare with binaries in  the field.  Early finding that the binary
frequency in  the Taurus-Auriga group is substantially  higher than in
the field  provided a strong  stimulus to these surveys.   The current
status  is  reviewed  by  \citet{Duchene2013}.   Statistics  of  young
binaries  contribute to  our understanding  of binary  star formation
and, consequently, of star formation in general.

A  popular  explanation  of  the  large  multiplicity  in  low-density
environments like Taurus (compared  to clusters and the field) invokes
dynamical  disruption  of wide  binaries  by  passing  stars. In  this
paradigm,  the binary  properties  at birth  are  universal, and  this
hypothetical primordial  binary population is  modified by ``dynamical
processing''  in clusters  \citep[e.g.][]{Kroupa2011}.   However, pure
dynamical evolution does  not explain the excess of  tight binaries in
Taurus-Auriga because  those pairs are  too ``hard''  (i.e. close
  and    strongly   bound)    for    being   destroyed    dynamically
\citep{King2012,Parker2014}.     Meanwhile,    large    hydrodynamical
simulations  of   collapsing  molecular  clouds   by  \citet{Bate2014}
demonstrated  that wide  binaries cannot  form in  dense environments,
while  the  binaries that  do  form are  hard  enough  to survive  the
$N$-body interactions  with their neighbors.  The idea  of a universal
primordial binary population and its dynamical processing is therefore
inconsistent with both theory and observations \citep{Duchene2018}.

Given that the binary formation  {\it does} depend on the environment,
it  is   of the utmost importance    to  characterize  it  observationally  in
star-forming regions of varying stellar density, metallicity, and age,
over a  large range of masses.   However, until very  recently we 
lacked statistically  significant samples of young stars  in the solar
vicinity spanning a wide range  of binary separations,  stellar masses
and  environments.   First and  foremost,  large, reliable  membership
samples are needed, and these have been hard to assemble, even for the
nearest  star-forming  regions,  which  are those  most  amenable  for
probing binarity down to small separations.  Unfortunately, even then,
the  small size  of  the stellar  population  of some  of the  closest
clusters and  associations \citep[e.g. $\epsilon$ Cha,][]{Briceno2017}
precludes  detailed statistical inferences. Therefore, our current  knowledge is
fragmentary and inconclusive \citep{Duchene2013,Reipurth2013}.

The nearby ($\sim$140\,pc) Upper Scorpius (USco) OB association (denser
part of  the larger Sco OB2  group) contains an updated  list of about
1600 known members \citep{L18}.   The association is highly structured
spatially,  rather than  expanding from  a common  center; its  age is
about 8 Myrs  \citep{Wright2018,David2019}. USco contains the largest
stellar  population  younger  than   $\sim  10$  Myr  within  300\,pc,
providing  an   unique  opportunity  to  learn  new   details  of  the
binary-star statistics.

Statistics  of binary  stars in  USco were  studied by  several groups
using  high-resolution  and   seeing-limited  imaging.   Yet,  only  a
fraction of known association members were examined so far.  Recently,
we studied multi-periodic stars in USco discovered by {\it Kepler} K2, assumed
to  be binaries  \citep[][hereafter  RSC18]{Rebull2018}, and,  indeed,
resolved most of them \citep{TokBri2018}.  The majority of those pairs
have not  been known previously  owing to the incompleteness  of prior
surveys.   We  found an  unusually  narrow  distribution of  projected
separations, with only a few binaries being wider than 1\arcsec.  The
distribution  differs markedly  from  the solar-type  binaries in  the
field  \citep{R10,FG67}. This  study convinced  us  that a  new large  and
uniform binary  survey of  USco is necessary.   The capability  of our
speckle instrument to observe hundreds of stars per night and the
astrometry  from   {\it  Gaia}   \citep{Gaia}  make the  present
survey both practical and timely.

The  input sample  of USco  members  is defined  and characterized  in
Section~\ref{sec:sample}. New  speckle-interferometric observations of
the  complete  sample  are  presented in  Section~\ref{sec:soar}.   In
Section~\ref{sec:bin} we  add data from {\it Gaia}  and other sources
and give  a comprehensive census  of resolved binaries.  The 
binary  statistics (distributions  of  mass ratio  and separation  and
their dependence  of mass)  are studied in  Section~\ref{sec:stat}. The
results   are   discussed   and   compared   to   other   studies   in
Section~\ref{sec:disc}.

\section{The sample of USco stars}
\label{sec:sample}

A  sample of the  members of  Upper Scorpius  association that  is not
biased with respect to the  binary population is the starting point of
our survey. At a first glance, the availability of accurate parallaxes
and proper motions  (PMs) in the {\it Gaia}  DR2 catalog \citep{Gaia},
hereafter  GDR2, makes  this task  easy.  However,  the GDR2  does not
provide  astrometry of  many resolved  binaries with  separations from
0\farcs1  to  $\sim$1\arcsec.  Moreover,  the  PMs  and parallaxes  of
binary  stars   in  GDR2  are  affected  by   their  orbital  motions.
Therefore,  a sample  based  on  the {\it  Gaia}  astrometry would  be
strongly biased  against binaries.  Recently  \citet{Damiani2019} used
GDR2 to study  membership of the Sco OB2  association, including USco.
Their work gives interesting insights on the spatial and kinematical
structure of  this region,  but it  is not a  good starting  point for
binary  statistics.   

We considered  the sample  of USco members  compiled by RSC18.   It is
restricted to objects within the  {\it Kepler} K2 Campaign~2 field and
contains  $\sim$1300 stars.   The  membership criteria  used by  these
authors  did not  rely  on the  GDR2  catalog that  had  not yet  been
released at the time of their study.  Comparison with the GDR2 reveals
that  the  RSC18 sample  of  USco  has  a non-negligible  fraction  of
non-members (about 15\%).  Moreover, we suspect that binary stars with
separations  of   a  few  arcseconds  and   components  of  comparable
brightness, semi-resolved  by {\it Kepler}, have been  removed from the
sample  because they  are not  suitable for  precise  photometry.  

We constructed a control sample  of USco members by selecting all GDR2
sources  with parallaxes  above 5\,mas  in the  rectangular  area with
$239\degr <  \alpha <  251\degr$ and $-30\degr  < \delta  < -17\degr$,
  filtering   on  parallax  (between   5.5  and  9  mas)   and  PM
  ($\mu^*_\alpha$ from $-$15 to $-$5, $\mu_\delta$ from $-$28 to $-$15
  mas~yr$^{-1}$),   and  keeping stars  brighter than  $G =  15$ mag.
Although the  control sample  of 664 targets  is biased  against close
binaries  and  is not  used here  for binary  statistics,  it is
useful for checking other biases. 

For our binary survey we use the large sample of USco members featured
in the  papers by \citet{L18}  and \citet{E18}, hereafter L18  and E18
(in  short,   the  Luhman's   sample).   These  authors   combine  the
astrometric membership  criteria (with suitable  allowance for errors)
with  photometry  and use the youth criteria  such as lithium line,
emissions,  and  IR excess.   The  highly  extincted  region near  the
$\rho$~Oph  cluster is explicitly  avoided.  Table~1  in L18  has 1631
entries.  The  matching Table~6  in E18 contains  1608 stars.   We use
here  the latter source,  as it  duplicates the  essential information
from L18 and contains, additionally, the $K$-band photometry.

The sample derived from the Table~6 of E18 has been cross-matched with
GDR2; all sources within 20\arcsec  radius (2800 au at 140\,pc) 
from each star were retrieved to identify potential binary companions.
The radius is chosen to avoid confusion between true binaries and
  random       pairs      of       association       members      (see
  Section~\ref{sec:cluster}).   Only 11  faint stars with $K \sim 15$
mag from the Luhman's sample are  not found in GDR2; they are probably
too faint in the {\it Gaia} $G$ band. Naturally, we do not discard the
107  stars  without parallaxes  in  GDR2  because  those are  resolved
binaries, as  we show below.  The  {\it Gaia} photometry  allows us to
compute the  $V$-band magnitude from  the $G$ magnitude  and  the color
index  $C$=BP$-$RP using the recommended transformation,\footnote{See  Chapter  5.3.7  of  {\it  Gaia}  DR2
  documentation                                                      at
  \url{https://gea.esac.esa.int/archive/documentation/GDR2/}. }
\begin{equation}
V \approx G +  0.0176 + 0.00686\; C +  0.1732 \; C^2 .
\label{eq:Vmag}
\end{equation}

For the survey,  we selected initially 744 stars  of spectral type M3V
or  earlier  because  the  remaining  stars  are  too  faint  for  our
instrument (see  below). We observed  in the $I$ filter  and therefore
estimated the $I$  magnitudes from $V$ and $V-K$  using an approximate
polynomial   relation  derived   by  fitting   isochrones   of  normal
main-sequence stars,
\begin{equation}
I \approx V -  0.196 + 0.2548\; (V-K) +  0.04567 \;(V - K)^2 .
\label{eq:Imag}
\end{equation}
Stars fainter than $I=13$ mag were removed from the sample, as well as
stars with GDR2  parallaxes less than 5\,mas and  larger than 15\,mas,
and  with a  high extinction  of  $A_K >  0.3$ mag.   This leaves  the
filtered sample of 614 stars  for our survey.  The cutoffs on spectral
type  and magnitude  retain  in  the sample  stars  more massive  than
$\sim$0.4 $M_\odot$. Ten stars  in the sample are secondary companions
to  other brighter  targets, so  the  sample consists  of 604  stellar
systems.

\begin{deluxetable}{ l l  l l }
\tabletypesize{\scriptsize}
\tablewidth{0pt}
\tablecaption{Filtered Luhman's sample 
\label{tab:sample}}
\tablehead{
\colhead{Col.} &
\colhead{Label} &
\colhead{Format} &
\colhead{Description, units} 
}
\startdata
1  & USn & I4 & Number in E18 \\
2  & $\alpha$ & F10.5 & R.A. (J2000), deg \\
3  & $\delta$ & F10.5 & Declination (J2000), deg \\
4  & $\varpi$ & F8.2  & Parallax, mas    \\
5  &  $\mu_\alpha^*$ & F8.2  & PM in R.A., mas~yr$^{-1}$    \\
6  & $\mu_\delta$    &  F8.2  & PM in Dec., mas~yr$^{-1}$    \\
7  & $V$ & F6.2  & $V$ magnitude, mag \\
8  &  $G$& F6.2  & Gaia magnitude, mag \\
9  &  $V-K$ & F6.2  & Color index, mag \\
10 & $A_K$ & F6.2  & Extinction from E18, mag \\ 
11 & $M_*$ & F6.2  & Estimated mass, $M_\odot$ \\
12 & SF & I2    & Secondary component
\enddata
\end{deluxetable}

\begin{figure}
\epsscale{1.1}
\plotone{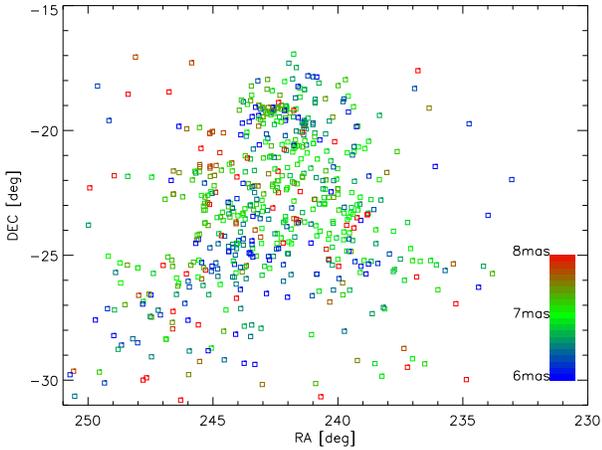}
\caption{Distribution of the filtered  sample on the sky.  The nominal
  limits  of  the Luhman's  sample  are  from  233\fdg75 to  251\fdg25
  in R.A. and  from $-$30\degr  to $-$16\degr  ~in declination.
  The region  near $\rho$~Oph is  avoided. The symbols are  colored by
  parallax (6\,mas in blue, 7\,mas in green, 8\,mas in red)  as
  shown by the colorbar.
\label{fig:sky}
}
\end{figure}

The filtered input sample  is presented in Table~\ref{tab:sample} (the
Table is available in full  electronically, its format is given in the
text).  In  column 12 we  assign a ``1''  to stars that  are secondary
companions, while  primary stars have  a ``0''.  The distribution  of the
stars  in  the  sky  in  Figure~\ref{fig:sky}  resembles  Figure~1  of
\citet{L18}.  The distribution of PMs, not plotted here, shows a tight
concentration to  the median  PM of $(-11.20,  -23.60)$ mas~yr$^{-1}$.
Luhman et al.   used the radius of 10  mas~yr$^{-1}$ around the median
PM for  selecting members  of USco, but  included some PM  outliers if
their  young age was  proven by  other criteria.   We note  a positive
correlation between  $\mu_\alpha$ and R.A.,  apparently reflecting the
complex       structure       of       the      USco       association
\citep{Wright2018,Damiani2019}.   The slope  is  $\sim$1 mas~yr$^{-1}$
per degree  of R.A.  Its inverse  value, sort of an  expansion age, is
4\,Myr.  No  other correlations between  position and PM  are evident.
The  median parallax is  6.99\,mas; 91\%  of parallaxes  are comprised
between 6 and 8\,mas.

The filtered sample contains 35  targets without parallaxes and PMs in
GDR2.   All those  stars,  without exception,  are resolved  binaries.
Whenever GDR2 does provide astrometry  for a binary, the results might
still be inaccurate and/or biased.  For example, all four targets with
parallaxes larger  than 10\,mas  (US0288, US0690, US0733,  and US0745)
are  binaries,  and their  parallax  errors  are  large, from  0.4  to
1.3\,mas.  Selecting  an input USco  sample based on  GDR2 astrometry,
like our control sample, inevitably creates a bias against binaries.

\begin{figure}
\epsscale{1.1}
\plotone{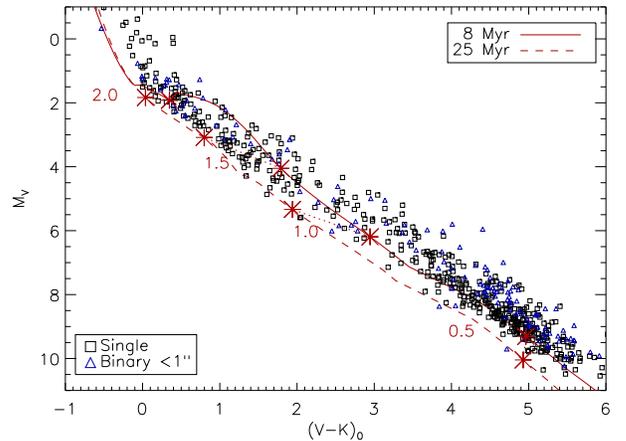}
\caption{Color-magnitude diagram for the filtered main sample.  The full
  and dashed lines are the 8 Myr and 25 Myr isochrones \citep{PARSEC}.
  Large  asterisks  and  numbers   mark  masses  on  both  isochrones.
  Individual  parallaxes and  extinctions are  used.  Binary stars
  closer than 1\arcsec ~are plotted by blue triangles.
\label{fig:CMD}
}
\end{figure}

The relation between the  extinction-corrected $(V-K)_0$ color and the
spectral  type is  almost linear  for  stars earlier  than K0V.   This
allows us to compute the intrinsic colors and the extinction for these
stars  from  the  spectral   types  provided  in  L18.   The  standard
extinction law implies  $E_{V-K} \approx 8 A_K$.  However,  we found a
shallower empirical relation $E_{V-K} \approx  5 A_K$.  We do not know
whether  this  discrepancy  is   caused  by  the  non-standard  (grey)
extinction in  USco or by  a bias of  the $A_K$ estimates in  L18.  We
apply the  extinction corrections using  the empirical slope of  5 and
obtain a  tighter sequence on the color-magnitude  diagram (CMD).  The
CMD  of the  main  sample is  shown  in Figure~\ref{fig:CMD}  (unknown
distances are assumed to  be 140\,pc, parallax 7.1\,mas).  The overall
band is quite thick for several reasons, e.g. binaries, an age spread,
or an intrinsic spread in luminosity. 

Candidate members of  USco from Table~2 of L18  have been evaluated in
the same way  as the main sample and cross-matched  with GDR2.  L18 do
not  provide extinction  for  these  stars.  For  the  most part,  the
candidates are  low-mass stars, already  well represented in  the main
sample.  Many candidates have spectral types later than our limit M3V.
For this reason we do not consider candidate members in our survey and
base it only on the filtered main sample.

\begin{figure}
\epsscale{1.1}
\plotone{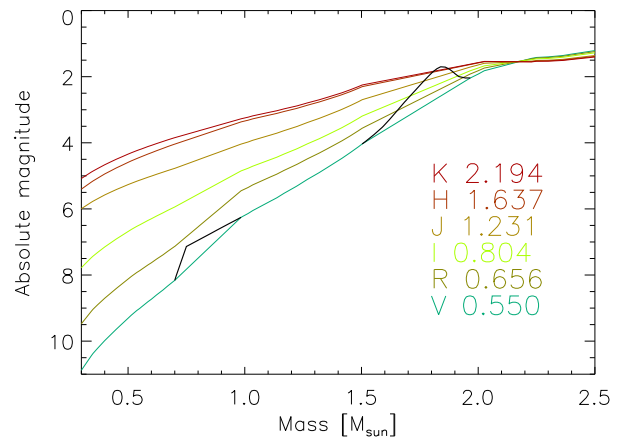}
\caption{Relation  between mass  and absolute  magnitude  in different
  bands from $V$ to $K$ (bottom  to top) according to the PARSEC 8-Myr
  isochrone  for solar metallicity  \citep{PARSEC}. Isochrones  in the
  mass  intervals  [1.5,  1.95]  \msun  ~and [0.7,  1.0]  \msun  ~were
  linearly  interpolated  to  avoid  the non-monotonic  behavior  (the
  original isochrone  in $V$ in  these areas is over-plotted  by thick
  black lines).
\label{fig:isochrone}
}
\end{figure}

\begin{figure}
\epsscale{1.1}
\plotone{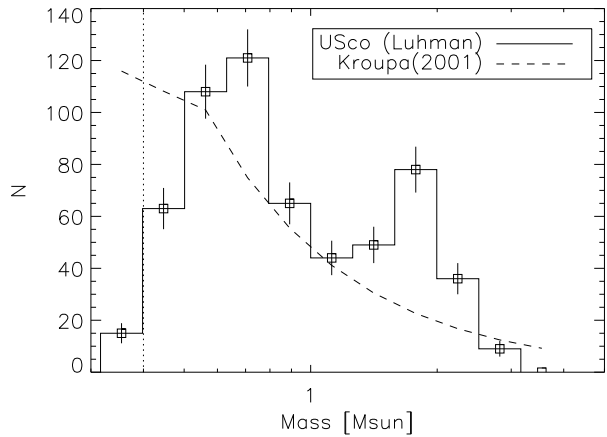}
\caption{Distribution of estimated masses in our filtered sample.  The
  dashed line  is the  \citet{Kroupa2001} initial mass  function (IMF)
  normalized to match the total  number of stars.  The decline at low mass
  is a  combination of removing  faint stars from the  filtered sample
  and its intrinsic incompleteness.  The  cutoff at 0.4 \msun ~is shown
  by the vertical line. 
\label{fig:masshist}
}
\end{figure}

We estimate masses  of the stars from their  absolute magnitudes $M_V$
corrected for  extinction using the  8 Myr PARSEC isochrone  for solar
metallicity  \citep{PARSEC}.   We  prefer  the $V$  band  because  the
dependence on  mass is steeper than  in the $K$ band,  hence errors in
$M_V$ (e.g.  caused  by binary companions) have less  influence on the
derived  masses  (Figure~\ref{fig:isochrone}).   In  order  to  get  a
monotonic relation between $M_V$  and mass, we eliminated the ``kink''
around 1.8  \msun ~and the  discontinuity around 0.7 \msun  ~by linear
interpolation of  all isochrones in  these two regions (see  the black
segments  in   Figure~\ref{fig:isochrone}).   Such  patching   of  the
isochrones  is necessary  for correct  evaluation of  masses  and mass
ratios.

Understandably, the  masses $M_*$ are  only crude and  possibly biased
estimates,  considering  the uncertainty  of  the  isochrones and  the
likely  spread  of  ages.   These  inaccurate masses  serve  only  for
relative  ranking  of  stars.   The  mass ratios  are  estimated  from
relative photometry  using patched  isochrones.  Here, only  the local
slope  of the  isochrones matters,  hence  the mass  ratios are  known
better than  the absolute masses. Masses $M_*$  assigned to unresolved
binary stars based  on their combined $M_V$ are,  on average, slightly
larger than  the masses of  their primary components (see  below).  In
the following,  we rank  all objects, single  and binary  alike, using
only $M_*$.

The  distribution of  points along  the main  sequence in  the  CMD in
Figure~\ref{fig:CMD} is  non-uniform, with less stars around
$M_V  \sim   6$.   The   distribution  of  absolute   magnitudes  and,
correspondingly,  masses  is  non-monotonic,  with  a  deficit  around
$\sim$1  \msun  (Figure~\ref{fig:masshist}).   This deficit  is  also
apparent in the distribution of raw magnitudes.  Luhman (2018, private
communication)  suggested that  his  sample is  incomplete for  G-type
stars.  Indeed, it is difficult to distinguish young G-type stars from
the  background  using standard  criteria  of  youth,  which are  more
reliable for  later spectral types. We  noted a similar  effect in the
RSC18 sample  of USco  members.  However, our  control sample  of USco
members based on  the GDR2 has a smooth  distribution of both absolute
magnitudes    and   derived   masses.     The   CMDs    presented   by
\citet{Damiani2019}  also appear  to have  a uniformly  populated main
sequence  without  gaps.   \citet{Luhman2018b} discussed  an  apparent
excess of late K-type stars in Taurus relative to the standard initial
mass function (IMF) and concluded that it is not real.  Therefore, the
relative deficit  of solar-mass stars  in the Luhman's sample  of USco
members  is likely  caused  by their  method  of candidate  selection.
Investigation of this effect is beyond the scope of our work.

\begin{figure}
\epsscale{1.1}
\plotone{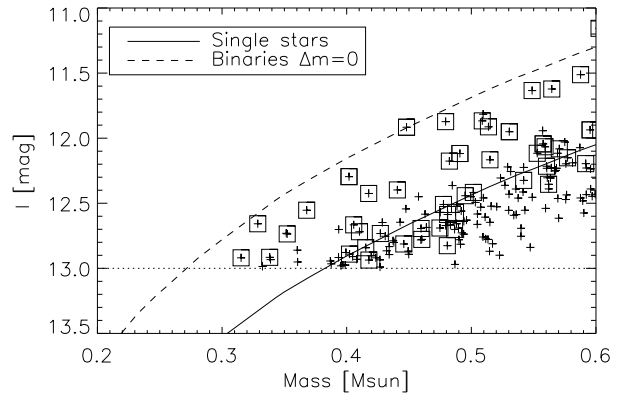}
\caption{Relation  between mass  and  $I$ magnitude  according to  the
  PARSEC 8-Myr  isochrone (line), assuming a distance  of 140\,pc. The
  magnitude cutoff  at $I=13$  mag is shown  by the  dotted horizontal
  line,  while  the  dashed  line  corresponds to  the  isochrone  for
  binaries with equal components, 0.75 mag brighter compared to single
  stars.  Estimated  masses and  $I$  magnitudes  of  the targets  are
  plotted as crosses, those of resolved binaries are marked by squares.
\label{fig:maglimit}
}
\end{figure}

The magnitude cutoff  $I < 13$ mag in  the filtered sample corresponds
to a single star of 0.38 \msun ~mass at a distance of 140\,pc, or $V =
15.9$ mag,  $V-I = 2.8$ mag, $V-K  = 5.3$ mag according  to the PARSEC
isochrone.  However, a  binary star is brighter than  a single star by
up  to  0.75 mag  and  may  be included  in  the  sample  even if  its
individual  components   are  below  the   photometric  cutoff.   This
situation, illustrated in Figure~\ref{fig:maglimit}, creates a bias in
favor of low-mass binaries (the Branch bias).  Indeed, we note that five
stars with $M_*  < 0.4$\msun ~are resolved binaries;  the remaining stars
in this  region could be tight  unresolved pairs, too.   However, at a
given $I$  magnitude binaries are  redder than single stars  and their
masses, estimated  from $M_V$, are below 0.4\,\msun.   By limiting our
statistical  analysis  to  stars  with  $M_*  >  0.4$\msun,  we  avoid
the Branch bias caused by the magnitude cutoff imposed on our sample.

\section{The SOAR  survey of USco}
\label{sec:soar}

\subsection{Instrument and data processing}

We  used the  high-resolution camera  (HRCam)  on the  4.1 m  Southern
Astrophysical Research  Telescope (SOAR) located at  Cerro Pach\'on in
Chile.  The  instrument, observing  procedure, and data  reduction are
covered  in \citep{HRCAM}; see  the recent  results and  references in
\citep{SAM18,SAM19}.

We applied twice for observing time to execute this survey through the
NOAO TAC,  but the time was not  granted.  So, we used  for this study
parts  of the  engineering nights  remaining after  completion  of the
technical work (mostly morning hours) and a fraction of time allocated
to  other  speckle  programs,  remaining  as a  result  of  our  highly
efficient observing procedure.  The observations started in 2018 March
\citep[these data are partly published in][]{TokBri2018} and continued
in  2019. Overall, we  used about  two nights  of telescope  time. Our
strategy   is  to   observe  all   targets,  regardless   of  prior
multiplicity surveys. 

The survey  has been  done in the  $I$ filter of  HRCam (824/170\,nm).
For each target, two data cubes of 400 frames each were recorded using
the   2$\times$2  binning  (effective   pixel  scale   31.5\,mas)  and
200$\times$200  binned pixels  region  of interest  (6\farcs30 on  the
sky). The  exposure time of  25\,ms  was  used for most  of the
targets; it  was increased to 50\,ms  and even to  100\,ms for fainter
stars and/or under worse seeing  conditions. Some data were also taken
without binning in a smaller 3\farcs15 field.

The data  cubes were  processed by the  SOAR speckle  pipeline, jointly
with other HRCam data \citep{HRCAM}. Companions are detected by
visual  inspection of  the speckle  auto-correlation  functions (ACFs)
and/or the speckle power spectra, where binary stars are manifested by
fringes. Parameters of binary stars (separation $\rho$, position angle
$\theta$,  and magnitude  difference  $\Delta m$)  and their  internal
errors are determined by fitting  the power spectra to the binary-star
model. Two (or more) data cubes of the same target give consistent
results in terms of binary-star astrometry and relative photometry,
while mutual agreement between the cubes provides another estimate of
the internal errors. The pixel scale and orientation are calibrated
using a set of wide binaries with well-modeled motion, observed in
each run together with the main programs. 

\begin{figure}
\epsscale{1.1}
\plotone{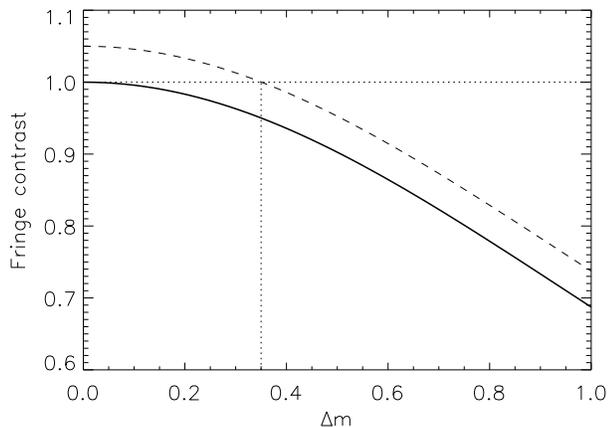}
\caption{Dependence of the fringe contrast in the speckle power
  spectrum on the magnitude difference of the binary star $\Delta
  m$. The dashed curve corresponds to a contrast error of +5\%.
\label{fig:fringe}
}
\end{figure}

Small  magnitude differences  $\Delta m  < 0.4$  mag are  not measured
reliably     by    speckle     interferometry.      As    shown     in
Figure~\ref{fig:fringe}, in  this regime the contrast  of fringes from
which $\Delta  m$ is calculated  depends on $\Delta  m$ quadratically.
An  error of  the measured  contrast by  +5\%, caused  by noise  or bias,
results in  $\Delta m =0$  assigned to all  binaries with $\Delta  m <
0.35$ mag.  This effect produces an  excess of binaries with $\Delta m =0$
in  the SOAR data.   Hence the  large mass  ratios $q>0.95$  cannot be
reliably  determined from  the relative  speckle photometry.   We take
this bias into account in our statistical analysis.

\begin{figure}
\epsscale{1.1}
\plotone{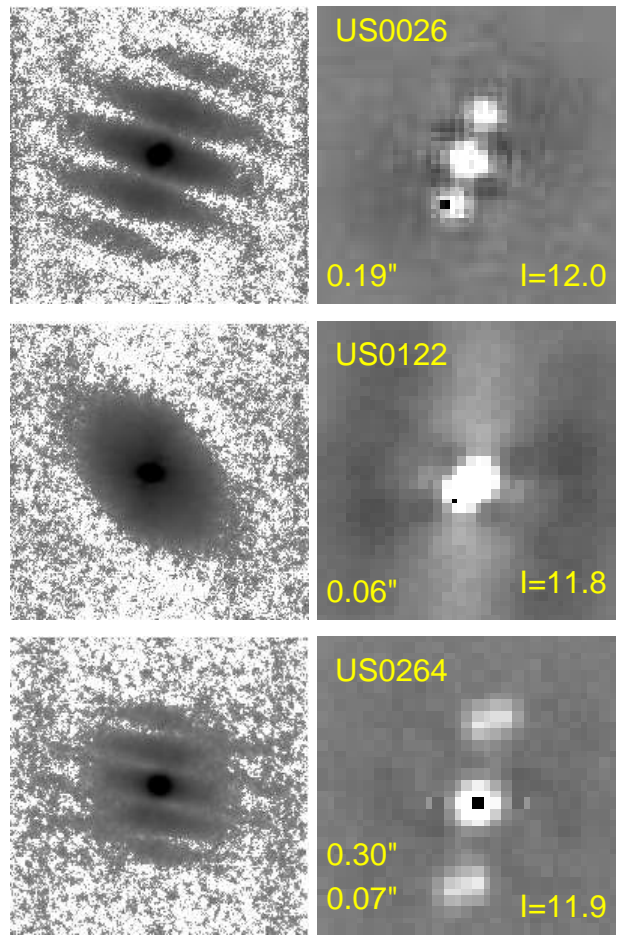}
\caption{Examples of three new  systems discovered in this survey. For
  each star, identified by its  number, the power spectrum is shown on
  the left (with a negative  logarithmic intensity scale), and the ACF
  on  the right,  in arbitrary  scale  (companions are marked  by   black
  dots).  The  separations  and  $I$-band  magnitudes  are  indicated.   The
  1\farcs58 companion to US0122 (KOH~55)  is not confirmed by SOAR and
  {\it Gaia}.  US0264 is a triple system of A,BC and B,C architecture.
\label{fig:images}
}
\end{figure}

Figure~\ref{fig:images}  illustrates  some  close  binaries   in  USco
discovered at  SOAR. Wide  companions are better  spotted in  the ACF,
while  companions near the  diffraction limit  are better  detected by
the elongation of the power spectrum.

\begin{figure}
\epsscale{1.1}
\plotone{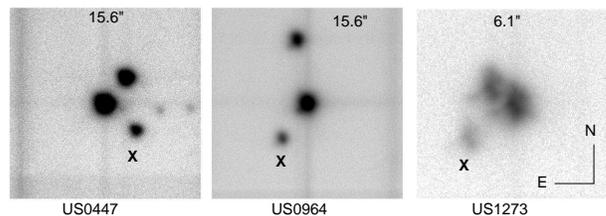}
\caption{Seeing-limited  images  of  three non-hierarchical  asterisms
  found at SOAR. Crosses mark optical companions with mismatching {\em
    Gaia} parallaxes. The angular size of each image is indicated.
\label{fig:trapezia}
}
\end{figure}

Three   targets   (US0447,   US0964,   and   US1273)   are   found   in
non-hierarchical configurations (trapezia) with comparable separations
of      a     few      arcseconds      between     the      components
(Figure~\ref{fig:trapezia}).  For  the first  two, images in  the full
15\farcs6 field were taken.  The GDR2 astrometry indicates that in all cases one
of  the  two  bright  companions  is  optical and  the  other  one  is
physical; optical  companions are marked  in Figure~\ref{fig:trapezia}
by  crosses.   Two  additional  faint  stars near  US0447  are  likely
optical,  too. USco is  close to  the Galactic  equator, where  a high
density of background stars can produce close asterisms.

\subsection{Detection limits of HRCam}

Detection  of binary  companions in  the speckle  ACF depends  on its
fluctuations  $\sigma$;   companions  above  $5\sigma$   are  reliably
detected, as  demonstrated by simulations of  artificial companions in
\citet{TMH10}. However, the $5\sigma$ criterion does not work for very
close pairs with separations below $\sim$0\farcs1, detected as fringes
in the power spectrum rather than peaks in the ACF.  The resolution is
determined by the diffraction  limit $\lambda/D$ (41.5\,mas in the $I$
filter),  but for  faint binaries  it can  be worse,  depending  on the
highest spatial  frequency where the  signal in the power  spectrum is
above the  noise level.  

\begin{figure}
\epsscale{1.1}
\plotone{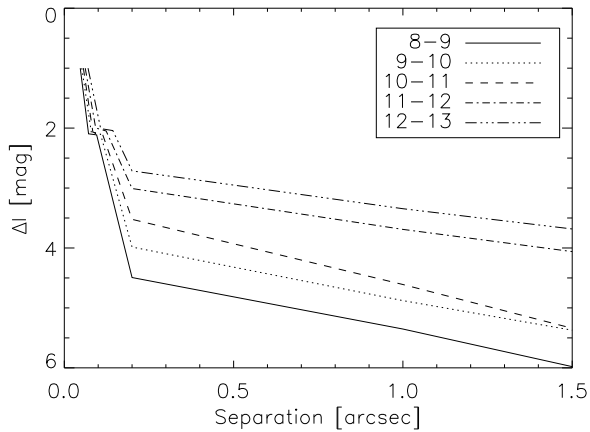}
\caption{Detection  limits for  the  speckle survey  of USco.   Median
  limiting contrast  $\Delta I$ at  six fixed separations  for targets
  grouped by  their $I$  magnitude is plotted.  Note that  the minimum
  separation also depends on the magnitude.  Separations from
    0\farcs05 to 1\farcs5 project to 7 and 210 au at 140\,pc
    distance. 
\label{fig:det}
}
\end{figure}

Reliable  knowledge  of  the  detection  limits  is  essential  for  a
binary-star survey like this  one.  Therefore, we studied detection of
binaries by simulating close companions using the actual power spectra
of faint  single stars with  artificial binary fringes. We  found that
the  effective  resolution  limit  corresponds to  $\rho_{\rm  min}  =
1/f_{\rm max}$, where $f_{\rm max} \le f_c = D/\lambda$ is the maximum
spatial  frequency containing  speckle signal  above  noise. Simulated
binaries with a  separation $\rho_{\rm min}$ and $\Delta  m \le 1$ mag
are detectable.  For wider binaries, the  standard $5\sigma$ criterion
is confirmed.

We modified our speckle pipeline and computed $\rho_{\rm min}$ for all
data cubes. The maximum detectable magnitude difference depends on the
binary separation, star brightness and,  of course, on the seeing that
influences the  strength of the  speckle signal.  Figure~\ref{fig:det}
shows the median detection limits at six fixed separations, grouped by
the $I$  magnitude of  the targets. Note  that the  minimum spearation
(the first point) also depends on the target magnitude. Under typical
conditions, the resolution is lost noticeably at $I> 12.5$ mag,
setting the limit of our survey.

\begin{figure}
\epsscale{1.1}
\plotone{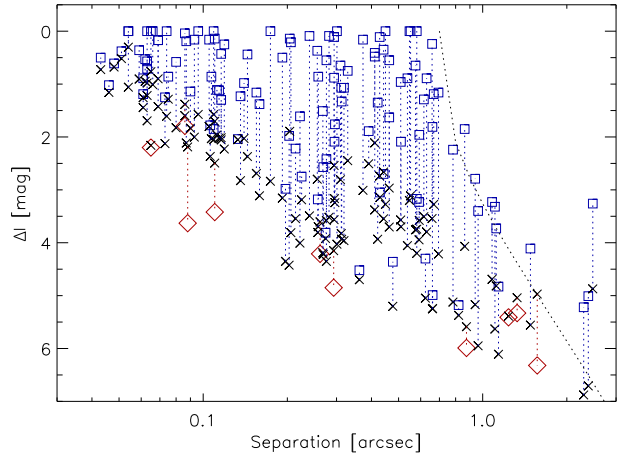}
\caption{The  magnitude  difference $\Delta  I$  of binary  companions
  measured  at  SOAR is  plotted  against  their  separation by  blue
  squares. They are connected by  dashed lines to the detection limits
  for these same stars (black crosses). A few cases where the companions are
  slightly fainter than the estimated detection limits are highlighted
  by red diamonds. Companions measured also by {\it Gaia} are removed
  from the plot, the adopted  {\it Gaia} detection limit is shown by the
  dotted line.
\label{fig:checkdet}
}
\end{figure}

The statistical  analysis below  uses the individual  detection limits
for each target.  They are selected as the best limits among available
data  cubes.   We checked  that  the  detected  binary companions  are
actually above the limits (Figure~\ref{fig:checkdet}). There are a few
exceptions where the companions  are fainter than the estimated limit.
However,  we should  bear in  mind that  the probability  of companion
detection is  a smooth function of  $\Delta m$ and $\rho$,  not a step
function.  The dotted line marks  the {\it Gaia} detection limit. Three
companions located in  this area are not found  in GDR2, but recovered
at SOAR.

\subsection{Table of SOAR results}

\begin{deluxetable}{ l l l l  }
\tabletypesize{\scriptsize}
\tablewidth{0pt}
\tablecaption{Results of SOAR observations
\label{tab:soar}}
\tablehead{
\colhead{Col.} &
\colhead{Label} &
\colhead{Format} &
\colhead{Description, units} 
}
\startdata
1  & WDS      &  A10 & WDS code (J2000) \\
2  & USn       & I4   & Number in E18 \\
3  & Name     &  A16 & Discoverer code or name \\   
4  & Date     & F8.3 & Date of observation, JY \\
5  & Filt.   &  A2   &  Filter  \\
6  & $\theta$ & F6.1 & Position angle, deg \\
7  & $\rho$   & F8.3 & Separation, arcsec \\
8  & $\sigma_\rho$ & F7.1 & Error on $\rho$, mas \\ 
9  & $\Delta m$    & F6.2 & Magnitude difference, mag \\
10 & $\rho_{\rm min}$ & F7.3 & Min. separation, arcsec \\ 
11 & $\Delta m_{0.15}$ & F7.2 & Max. $\Delta m$ at 0\farcs15, mag \\
12 & $\Delta m_{1}$   & F7.2 & Max. $\Delta m$ at 1\arcsec, mag 
\enddata
\end{deluxetable}

The results  of observations with HRCam  at SOAR are  presented in the
electronic Table~\ref{tab:soar}. Its columns contain: (1) the WDS code
based on  J2000 coordinates (for objects  that are not  present in the
WDS these  codes were  created); (2) the  source number USn  from E18,
same  as in Table~\ref{tab:sample};  (3) the  discoverer name  and, if
necessary,  component designation,  taken  from the  WDS for  resolved
known pairs or derived from  the USn numbers otherwise.  The following
columns give (4) Julian year  of observation, (5) filter, (6) position
angle, (7) separation, (8) error  of separation, and (9) the magnitude
difference.   For unresolved  sources all  these parameters  are zero.
The relative  photometry and astrometry of resolved  triples refers to
pairings  between individual  components as  indicated (e.g.  A,B and
B,C, but not  A,BC).   The  last  three columns  give  (10)  the  minimum
detectable  separation $\rho_{\rm min}$,  (11) the  maximum detectable
magnitude difference  at 0\farcs15, and  (12) same at  1\arcsec.  More
detailed  information is presented  in the  merged table  of detection
limits described below.

The  data assembled  in Table~\ref{tab:soar}  come from  a  variety of
observing programs  executed with HRCam. They  include observations of
multi-periodic stars published  by \citet{TokBri2018}, observations of
binaries  with  orbital motion,  etc.   We  omitted several  redundant
observations  taken before 2016.   Some stars  were visited  more than
once,  either because  they  belonged to  different  programs or  were
followed to  confirm the  discovery or to  detect the  orbital motion.
Overall,  Table~\ref{tab:soar} contains  706 rows,  with  187 resolved
pairs or subsystems; 89 resolutions are new  (34 of those are also
  found  in GDR2),  another 29  are published  by \citet{TokBri2018}.
Almost  half  of USco  binaries  known today  (118  out  of 250)  were
discovered at SOAR.  The total number of processed data cubes is 1536.
All 614 targets have been observed at least once.

Many  close  pairs  in  USco  discovered  here  have  short  estimated
periods. Their orbits  can be determined within a  few years, yielding
masses for testing evolutionary  tracks of young stars. Fast orbital motion
of some  pairs is   evidenced by repeated measurements  at SOAR
taken within a year.

\section{Binary stars in Upper Scorpius}
\label{sec:bin}

In this  Section, we combine the  results of our SOAR  survey with the
previously available data  and the information from GDR2  to produce a
catalog  of   250  physical  binary  stars  in   USco  for  subsequent
statistical analysis.

\subsection{Previous surveys} 

The statistical analysis presented below does not rely entirely on the
SOAR data and includes the results of previous multiplicity surveys in
USco. The binaries detected in these  surveys or known from the era of
visual observers are included in  the general list.  We also take into
account the number of targets and the detection limits of each survey.
The  detection  limits are  defined  here  as  the maximum  detectable
$\Delta  m_i$  at  six  separations $\rho_i$.   The  first  separation
corresponds to  the angular resolution $\rho_{\rm min}$,  and the last
to the maximum surveyed separation  (usually half of the imaging field
size).   The  detection   limits  are  linearly  interpolated  between
$\rho_i$ and,  knowing the imaging  wavelength, can be  converted into
the minimum detectable mass ratios $q_{\rm min}(\rho)$ for each of our
targets using  the isochrones. All  detection limits are  assembled in
Table~\ref{tab:det}.

\begin{figure}
\epsscale{1.1}
\plotone{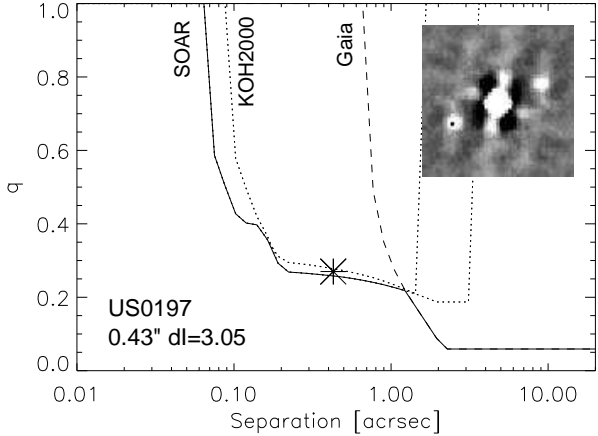}
\caption{Limits  of  companion  detection  around US0197.   The  faint
  companion at 0\farcs43 with $q=0.27$  is marked by a large asterisk;
  the corresponding ACF is shown in the insert, where the black dot marks the companion. Secondary spikes in the
  ACF are artifacts.  
\label{fig:det197}
}
\end{figure}

Figure~\ref{fig:det197} illustrates the  combined detection limits for
one of our  targets, US0197 ($M_* = 0.56$ \msun, $I  = 12.6$ mag).  It
has been observed at  high angular resolution by \citet{Koehler2000}
and at  SOAR; the  latter detected a  companion at $\rho  = 0\farcs43$
with $\Delta I = 3.05$ mag which, in principle, could be also found by
Koehler et  al. At  separations beyond 1\arcsec, {\it  Gaia} has  a deeper
detection limit compared to speckle imaging.

\begin{deluxetable}{ l l l l  }
\tabletypesize{\scriptsize}
\tablewidth{0pt}
\tablecaption{Detection limits
\label{tab:det}}
\tablehead{
\colhead{Col.} &
\colhead{Label} &
\colhead{Format} &
\colhead{Description, units} 
}
\startdata
1  & USn      & I4   & Number in E18 \\
2  & Ref     &  A8 & Reference\tablenotemark{a} \\   
3  & Meth    & A3 & Method\tablenotemark{b} \\
4 &  $\lambda$     &  F6.2      & Imaging wavelength, $\mu$m \\
5--10 &  $\rho_i$   &  F6.3      & Separations, arcsec \\
11--16 & $\Delta m_i$ & F6.2 & Maximum $\Delta m$, mag
\enddata
\tablenotetext{a}{See Table~\ref{tab:bib}.}
\tablenotetext{b}{Methods: 
AO -- adaptive optics;
SI -- speckle interferometry;
q -- mass ratios are given instead of $\Delta m$.
}
\end{deluxetable}


\begin{deluxetable*}{l l c c c l}[ht]
\tabletypesize{\scriptsize}    
\tablecaption{Multiplicity surveys of USco
\label{tab:bib} }                    
\tablewidth{0pt}     
\tablehead{ 
\colhead{Label}  &
\colhead{Reference}  &  
\colhead{$N$}  &  
\colhead{$\lambda$}  &  
\colhead{$\rho_{\rm min}$}  &  
\colhead{Method\tablenotemark{a}}  \\
&  & & 
\colhead{($\mu$m)}  &  
\colhead{(arcsec)}  &  
}
\startdata
KOH2000 & \citet{Koehler2000} & 41 & 2.2 & 0.06 & SI \\
KOU05 & \citet{KOU05} & 50 & 2.2 & 0.2 &  AO  \\
MH09  & \citet{MH09}  & 16 & 2.2 & 0.1 & AO  \\
KRS12 & \citet{Kraus2012} & 44 & 2.2 & var & AO, Seeing \\
LAF14 &  \citet{Laf2014} & 73  & 2.2 & 0.10 & AO \\
SOAR  & This work   &    614   & 0.8 & 0.05 & SI
\enddata
\tablenotetext{a}{Methods:
SI -- speckle interferometry;
AO -- adaptive optics; 
Seeing -- seeing-limited.}
\end{deluxetable*}


Table~\ref{tab:bib}  summarizes  in   compact  form  relevant  imaging
surveys  of USco,  in chronological  order.   It gives  the number  of
targets $N$ belonging to  our sample, the imaging wavelength $\lambda$
in microns, and the angular resolution $\rho_{\rm min}$ in arcseconds.
For   each  publication,  the   corresponding  electronic   table  was
retrieved, read by a specially written IDL program, cross-matched with
the  sample  by  coordinates,  and  exported in  the  standard  format
($\rho_i$  and $\Delta m_i$)  into a  text file.   When the  number of
detection limits given in the paper  is less than six, the last limits
are duplicated.  When the number is  larger than six, we use the first
five  closest  separations  and  the last  widest  separation.   Short
comments on each survey are provided  in the rest of this Section. The
present sample exceeds previous samples by an order of magnitude.

\citet{Koehler2000} have  performed the pioneering survey  of USco and
other young associations using  speckle interferometry in the $K$ band
at the ESO  3.5 m NTT telescope.  For all their  targets, we adopt the
fixed detection limits from Figure~3  of their paper: $\rho_i = [0.06,
  0.1, 0.13, 0.6, 1.0, 6.0]$ arcsec,  $\Delta m_i = [0.0, 2.5, 3.0, 4,
  4.5, 5.0]$ mag.

\citet{KOU05} surveyed relatively bright stars of spectral types B, A,
and F with adaptive optics (AO). They used the ADONIS AO system on the
3.6 m ESO telescope at La Silla.  All targets were observed in the $K$
band, some also  in the $J$ and  $H$ bands.  We use here  only the $K$
band  data  (they  are  deeper   in  the  mass  ratio,  but  lower  in
resolution).  Their  paper does  not provide the  individual detection
limits,  only the  summary plot  in their  Figure~3 with  an empirical
limiting  line.  We  reproduced  a  similar plot  from  their data  on
resolved  companions and adopted  the upper  envelope as  the relevant
detection limit: $\rho_i = [0.2, 0.5, 1, 2, 5, 10]$ arcsec and $\Delta
m  = [0.5,  2.7,  5, 7,  8, 8]$  mag.   The detection  depth at  large
separations does  not matter because we confirm  wide companions using
{\it Gaia}.

\citet{MH09} observed young stars,  including several USco members, in
the $K$  band using AO at  the Palomar and Keck  telescopes.  We adopt
fixed detection  limits, $\rho_i  = [0.09, 1.0,  2.0, 5.0,  5.0, 5.0]$
arcsec and $\Delta m_i = [0.0, 17.2, 19.4, 20.3, 20.3, 20.3]$ mag.

\citet{Kraus2012} published  a compilation of  multiplicity surveys in
USco  done  both  with   AO  and  at  the  seeing-limited  resolution.
Individual  detection limits  are  retrieved from  their Table~7  that
lists minimum mass  ratios vs.  projected separations in  au, for each
star  individually.    The  data  are  ingested   as  published,  with
separations converted  back to angular units.  We  assume that earlier
works  by this  group  are included  in  this compilation  and do  not
consult their earlier papers.


\citet{Laf2014} observed  91 stars in  USco with the AO  instrument at
the Gemini North  telescope. Their input sample was  based on the same
list of bright USco members as used by prior surveys, hence there is a
large number  of targets in  common with prior  work. We use  the data
from  their Tables 1  and 2  that provide  individual limits  for each
target. Lafreni\`ere et al.  noted that the multiplicity fraction does
not decline  with mass (unlike stars  in the field)  and that binaries
with  comparable  masses and  large  separations  are  rare. Below  we
confirm both their conslusions for our much larger sample.

Several publications are not considered in the table of detection
limits; they are commented briefly. 

\citet{ST02} observed 115  stars of spectral types O and  B in the Sco
OB2 association, including USco.  However, none belongs to our sample.

\citet{Bou06} observed low-mass members of USco and resolved some
multiple systems. Unfortunately, their paper does not contain the list
of all observed targets and the detection limits. None of these
binaries is present in our list of resolved pairs. 

\citet{Janson2013} observed   138  bright stars in  the Sco-Cen
region,  not covered  by prior  work, using  the NICI  AO  instrument at
Gemini South.   Examination of their  Table~1 reveals no  matches with
our sample.   Interestingly, they found a total  absence of relatively
wide  pairs with  small $\Delta  m$, an  effect likely  caused  by the
construction  of their  sample that  avoided known  binaries  from the
outset.

\citet{Hnk2015}   used  high-contrast   imaging  and   detected  faint
companions to  three stars in  USco, two of which,  HIP~78196 (US0193)
and HIP~79124  (US0708), belong to  our sample.  They do  not provide
the list of all observed targets.  

Our  sample contains  126 targets  with {\it  Hipparcos}  numbers. The
limits of  companion detection by  {\it Hipparcos} are adopted  in the
same way  as for the  67-pc sample of solar-type  stars \citep{FG67}:
$\rho_i =  [0.09, 0.14, 0.4, 1, 5,  10]$ arcseconds, $\Delta m_i  = [0, 2.2,
  4.0, 4.1, 4.2, 4.3]$ mag, wavelength 0.5 micron.

\subsection{Binaries in the Gaia DR2} 

\begin{figure}
\epsscale{1.1}
\plotone{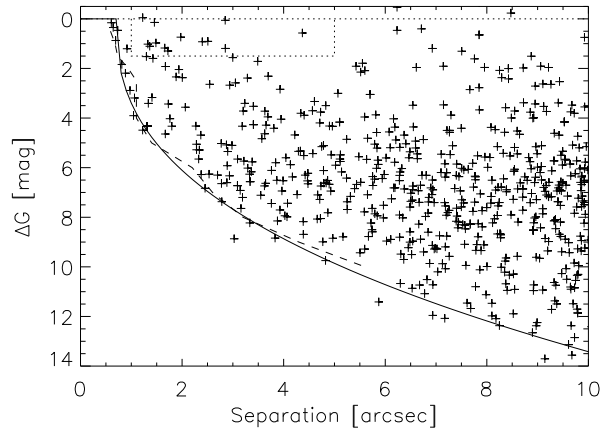}
\caption{Magnitude  difference  $\Delta  G$  vs.  separation  for  all
  companions in {\it Gaia} DR2  found near our targets.  The full line
  is  the  adopted detection  limit  according  to (\ref{eq:dg}),  the
  dashed line is the  50\% detection limit from \citet{Brandeker2019}.
  The dotted rectangle indicates the zone of blended targets.
\label{fig:gaia}
}
\end{figure}

As  mentioned above, all  sources within  20\arcsec ~from  our targets
were    retrieved     from    the    GDR2     catalog    \citep{Gaia}.
Figure~\ref{fig:gaia}  shows the  magnitude difference  vs. separation
for all {\it  Gaia} pairs.  The lower envelope  is approximated by the
formula
\begin{equation}
\Delta G (\rho) = 5.5(\rho- 0.7)^{0.4}, \;\;\; \rho > 0\farcs7,  
\label{eq:dg}
\end{equation}
which describes  the {\it Gaia}  detection limit.  Similar  limits for
companion    detection    in    {\it    Gaia}    were    derived    by
\citet{Brandeker2019}.   By companions we mean here  both the real
  (physical) binaries  and the random optical pairs,  according to the
  established double star  terminology.    In addition to  the contrast
limit,  stars  fainter  than  $G =  20.5$  mag are  not
present in GDR2,  reducing the number of pairs  with large $\Delta G$.
The GDR2 magnitude limit, relevant for faint targets, is accounted for
by our model of {\it Gaia} companion detection.

Naturally,  the majority  of companions  in  Figure~\ref{fig:gaia} are
unrelated  (optical)  stars.  The  average  number  of companions  per
target with $\Delta  G < 5$ mag and separation  less that $\rho$ grows
quadratically   as   $N_{\Delta G < 5}(\rho)   \approx   0.81   (   \rho
/20\arcsec)^2$.   According to this  formula, we  expect to  find five
optical companions  with $\Delta G <  5$ mag and $\rho  < 2$\arcsec ~in
the whole sample; the actual number of such {\it Gaia} pairs is 40.

We  consider  all  {\it  Gaia}   pairs  with  $\rho  <  2$\arcsec  ~as
potentially  physical. Some  (but not  all) of  these close  pairs are
classified as physical or optical because parallaxes of the companions
are  present in  GDR2.   Most  close {\it  Gaia}  companions are  also
detected at  SOAR and  confirmed as physical  (co-moving), with  a few
exceptions like US1353.  The wider  {\it Gaia} companions with $\rho >
2\arcsec$ are accepted as physical  only if they have astrometric data
in GDR2  that confirm their  membership in USco. Wide  companions that
are themselves  close pairs and  hence lack {\it Gaia}  astrometry are
excluded from our survey.

\begin{figure}
\epsscale{1.1}
\plotone{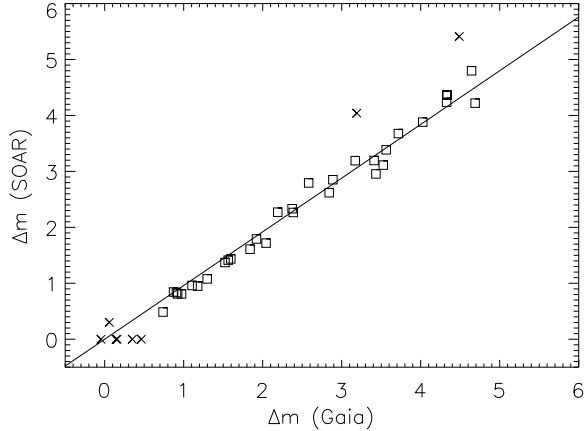}
\caption{Comparison between  relative photometry
  of  pairs measured by  {\it Gaia} and at  SOAR. The
 outliers  are marked by  crosses, and the  line is
  $\Delta I = 0.96 \Delta G$.
\label{fig:soargaia}
}
\end{figure}

There are 40 pairs wider than 0\farcs6 detected by both {\it Gaia} and
SOAR.  Their  angular separations match  very well. The  median offset
$\rho_{\rm SOAR}  - \rho_{Gaia}  $ is  9 mas, and  its rms  scatter is
10\,mas  after removing  the optical  pair US1353  (its  companion has
moved  by   63\,mas  between   2015.5  and  2019.2).    The  magnitude
differences are  also in good  agreement, with $\Delta I  \approx 0.96
\Delta  G$ (line  in Figure~\ref{fig:soargaia}).   However,  two pairs
(US0820  and US1081) plotted  by crosses  have a  substantially larger
$\Delta I$  measured at SOAR;  their separations are  around 1\arcsec.
The latter was observed in  a small 3\farcs15 field where the relative
photometry could be biased  by vignetting.  Photometric variability of
some companions (flares or dimmings) cannot be ruled out because these
stars are  young.  We note that  speckle photometry has  a tendency to
under-estimate small $\Delta  I <0.5$ mag by assigning  $\Delta I =0$,
as noted  above (see Figure~\ref{fig:fringe}).  We use  the {\it Gaia}
photometry when it is available.

Some wide  {\it Gaia} pairs consist  of two members of  our sample. In
this    case    the     secondary    components    are    marked    in
Table~\ref{tab:sample} and not counted as separate targets, with a few
exceptions like  trapezia discussed  below. Targets US0432  and US0613
are  paired to  USco  members that  are  not present  in the  Luhman's
sample, at separations 16\farcs2 and 19\farcs0 and brighter by 1.7 and
0.5 mag,  respectively.  These  pairs are not  included in  our binary
catalog.   However, the  companion  to US1248  at  8\farcs5, 0.23  mag
brighter, is included with $\Delta G = 0.23$ mag.

\subsection{Binaries in the WDS}
\label{sec:WDS}

We retrieved all known pairs  in our sample from the Washington Double
Star Catalog,  WDS \citep{WDS}. Only pairs  with $\rho<20$\arcsec ~and
$\Delta m  < 6$  mag are considered  to avoid numerous  optical pairs.
High-contrast imaging revealed many  faint companions around some USco
stars,  all  faithfully  recorded  in   the  WDS,  and  most  of  them
optical. We keep only the binaries proven to be physical by {\it Gaia}
and ignore  the rest.  Pre-{\it Gaia} multiplicity  surveys might have
included   some   optical   pairs   and   hence   over-estimated   the
multiplicity. The majority  of WDS binaries are also  detected at SOAR
and/or by  {\it Gaia}, and only  for 16 systems we  use the literature
data.  Most of those are  close pairs with a large contrast discovered
by aperture  masking and undetectable  at SOAR.  Their  parameters are
outside the range of separations  and mass ratios studied here, but we
keep  these pairs for  completeness.  A  few unconfirmed  WDS binaries
(e.g.  those discovered by lunar occultations) are omitted.

\subsection{Blended targets}
\label{sec:blend}

In  the  following,  we  note   the  paucity  of  USco  binaries  with
separations of a few arcseconds  and nearly equal components.  When we
first saw  this effect in the  sample of RSC18, there  was a suspicion
that  such stars were  removed from  the {\it  Kepler} K2  campaign as
unsuitable  for  precise   photometry.   The  Luhman's  sample,  being
independent of {\it Kepler}, still uses photometry and astrometry that
could be affected by the resolved nature of some stars, leading to the
rejection of  blended targets.   For example, semi-resolved  stars are
missing  from the  2MASS point  source catalog.   These stars  are not
suitable  as  reference for  AO  systems  and  could be  removed  from
AO-based  multiplicity surveys.   For brevity,  we call  binaries with
$1\arcsec  <  \rho  <  5\arcsec$  and  $\Delta  m  <  1.5$  mag  (this
corresponds roughly to $q> 0.7$)  {\it blended targets.}  To prove the
reality of the deficit of  wide binaries  with similar components
in USco,  we must verify that  the input sample is  not biased against
blended targets.

To address  this concern, we  selected from the  WDS all pairs  in the
area  covered  by  the   Luhman's  sample,  with  PMs  within  $\pm$10
mas~yr$^{-1}$  from  the  mean   PM  of  USco  and  separations  $\rho
<20\arcsec$.   The resulting  list was  cross-matched with  GDR2.  The
subset of  WDS pairs in the  blending regime was  examined manually to
reject  non-members of  USco (based  on parallax)  and  optical pairs,
leaving only  six physical pairs.  Three  of those are  present in our
binary catalog, one  is too faint, and two could  be indeed the missed
blended targets.   Their WDS codes are  16116$-$1839 and 16256$-$2327,
primary $G$  magnitudes 12.37,  5.54, mag, separations  1.3\arcsec and
3.0\arcsec, $\Delta m$ 0.50 and 0.67 mag. For consistency, we do not
add those targets to the Luhman's sample. 

\begin{figure}
\epsscale{1.1}
\plotone{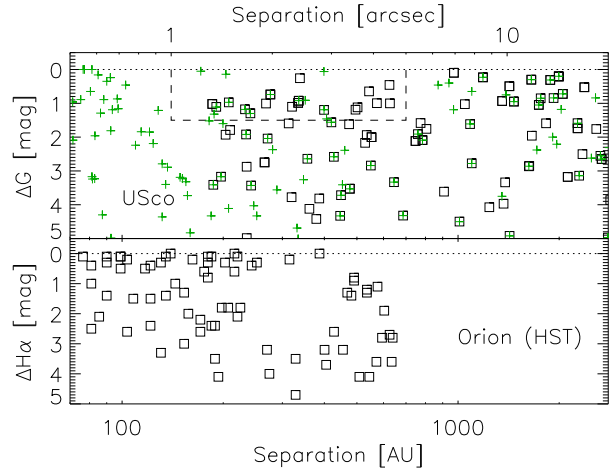}
\caption{Magnitude difference  vs.  separation.  The  upper plot shows
  pairs  of USco  members in  the control  GDR2 sample  (squares). For
  comparison,  pairs  in  our  binary  catalog are  plotted  by  green
  crosses.  The  regime of blended  targets is outlined by  the dashed
  rectangle.  The lower plot refers  to binaries in the Orion from the
  Table~1  of \citet{Reipurth2007},  discovered with  the  Hubble Space
  Telescope.
\label{fig:blend}
}
\end{figure}

The  {\it  Gaia}  census  of  stars offers  an  independent  check  of
potential  bias against  blended targets.  All such  pairs  are easily
detectable   by   {\it   Gaia}    (see   the   dotted   rectangle   in
Figure~\ref{fig:gaia}). Using  our control sample of  664 USco members
based exclusively on  GDR2, we found 18 pairs  in the blending regime.
They are  plotted in Figure~\ref{fig:blend}; 16 of  them are confirmed
as  physical by  {\it Gaia}  astrometry  of the  companions, two  have
unknown  status.  For comparison,  our  binary  catalog  based on  the
Luhman's  sample contains 13  blended targets  among 604  systems (one
optical pair  is removed from  the catalog). The numbers  and relative
frequencies of blended targets  in both samples are consistent, within
statistical errors.  We  also used the sample of  USco members derived
from the lists of \citet{Damiani2019} and reached the same conclusion.
The Luhman's  sample cannot miss more  than a few  blended targets, if
any.   We   infer  that  its   bias  against  blended   targets  is
insignificant and the paucity of such binaries is real.

The paucity of  near-equal binaries with separations of  a few hundred
au, compared  to smaller  and larger separations,  may exist  in other
star-forming regions,  for exampe in  the Orion Nebular  Cluster (ONC)
studied by \citet{Reipurth2007}.  These  authors noted a sharp decline
in the  binary frequency at  separations $>$225\,au and related  it to
the  dynamical  disruption of  wide  binaries  in  the dense  cluster.
However, 10 pairs out  of 78 in their Table~1 have $\Delta  m < 2$ mag
and  projected  separations  $s  >500$\,au  (see the  lower  panel  of
Figure~\ref{fig:blend}).   These binaries  are  likely physical  (wide
optical pairs  tend to have large $\Delta  m$).  Therefore, relatively
wide binaries with  comparable components exist in the  ONC as well as
in USco.  However, in both regions such pairs are rare at intermediate
separations from 200 to 500 au.

\subsection{Combined list of binary stars}
\label{sec:comp}

\begin{deluxetable}{ l l l l  }
\tabletypesize{\scriptsize}
\tablewidth{0pt}
\tablecaption{List of binaries in USco
\label{tab:comp}}
\tablehead{
\colhead{Col.} &
\colhead{Label} &
\colhead{Format} &
\colhead{Description, units} 
}
\startdata
1  & WDS      &  A10 & WDS code (J2000) \\
2  & USn      & I4   & Number in E18 \\
3  & Name     &  A16 & Discoverer code or name \\   
4  & Date     & F8.2 & Date of observation, year \\
5  & $\theta$ & F6.1 & Position angle, deg \\
6  & $\rho$   & F8.3 & Separation, arcsec \\
7  & $\Delta m$    & F6.2 & Magnitude difference, mag \\
8 &  $\lambda$     &  F6.2      & Wavelength, $\mu$m \\
9 &  $m_1$       &  F6.2      & Primary mass, \msun \\
10 & $q$         & F6.3 & Mass ratio \\
11 & $L$         & I2 & Hierarchical level
\enddata
\end{deluxetable}

The lists of companions from three sources, {\it Gaia}, SOAR, and WDS,
were  merged and examined  to produce  the final  list of  binaries in
Table~\ref{tab:comp}, selecting from multiple sources in this order of
preference.   The  first  three   columns  are  similar  to  those  of
Table~\ref{tab:soar}.   Then follow the date, the position angle  $\theta$, the
separation  $\rho$,  the  magnitude  difference $\Delta  m$,  and  the
wavelength $\lambda$  to which it  refers (0.6 $\mu$m for  {\it Gaia},
0.8 $\mu$m  for SOAR  in the  $I$ filter).  The  columns (9)  and (10)
contain the  estimated mass of the primary  component (contribution of
the secondary to  the combined light of close  pairs is accounted for)
and the mass ratio. The last  column (11) codes the hierarchy (level 1
for outer  systems, level 11  for inner subsystems around  primary, 12
for secondary subsystems).  We  estimated mass ratios from the patched
PARSEC 8\,Myr isochrones as follows.  The isochrone in the photometric
band  closest  to the  imaging  $\lambda$  is  selected. The  absolute
magnitude in this  band is computed from the  crude mass $M_*$ estimated
for this  target in the main  sample. The absolute  magnitudes of both
components are  computed from the measured $\Delta  m$, accounting for
the contribution of the secondary  component to the combined light for
close pairs unresolved by {\it Gaia} (remember that the $V$ magnitudes
are  derived from  the {\it  Gaia}  photometry).  The  masses of  both
components  are  interpolated  back   from  the  same  isochrone.   As
expected, for  some close  pairs the primary  mass $m_1$ is  less than
$M_*$,  but  the difference  is  typically  small.   The median  ratio
$m_1/M_*$ is 0.99, the minimum ratio is 0.7. We consistently use $M_*$
for ranking all stars in mass.

Seven subsystems belonging to  the secondary components of wider pairs
(level 12) are included in our analysis.  However, their mass $M_*$ is
set to  the secondary  mass $m_2$  of the outer  pair. Three  of those
subsystems  fall below  our cutoff  at 0.4  \msun, the  remaining four
contribute to the statistics at their masses, not at the masses of the
wide primary  components.  There  are also 11  subsystems of  level 11
belonging to the primary components of wider pairs. 

The list  contains 250 pairs,  counting all subsystems  separately.  A
large  number  of  physical  pairs,  70,  were  found  in  {\it  Gaia}
(discovery code  GAnnnn, where  nnnn is the  number in E18);   34 of
those are independently confirmed at SOAR.  We also used {\it Gaia} to
discard some optical  companions listed in the WDS.   When the WDS and
SOAR  pairs were  measured also  by {\it  Gaia}, the  latter  data are
preferred; they are distinguished by the date of 2015.5 and $\lambda =
0.6$\,$\mu$m. Overall,  102 pairs have {\it  Gaia} relative astrometry
and photometry. We provide additional comments on some binaries in the
Appendix.

The  catalog   includes  28  binaries   from  \citet{TokBri2018}  with
discoverer  codes TOK* assigned  to them  by the  WDS, and  another 55
pairs  resolved   at  SOAR  later  (names  starting   with  US).   All
measurements of  SOAR pairs  are listed in  Table~\ref{tab:soar}.  The
remaining  97 pairs  with  various discoverer  codes  were known  from
previous work.   Most of  them are confirmed  by {\it Gaia}  and/or by
SOAR.  We  prefer the SOAR measurements  over those given  in the WDS,
and give them  for 132 close pairs unresolved by  {\it Gaia}.  Only 16
pairs in Table~\ref{tab:comp}, unresolved by both {\it Gaia} and SOAR,
have relative astrometry and photometry retrieved from the literature.

\section{Binary statistics}
\label{sec:stat}

\begin{figure}
\epsscale{1.1}
\plotone{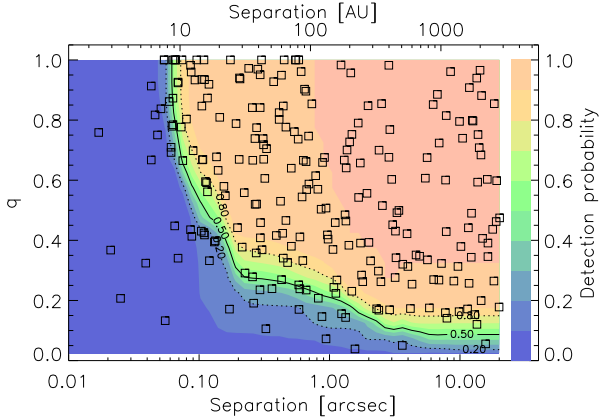}
\caption{Mass ratio vs. separation for all binaries in our sample. The
  colors represent  average detection probability  from 0 to  1 (color
  bar  and scale on  the right);  its contours  at 0.2, 0.5,  and 0.8  levels are
  overplotted. 
\label{fig:stat}
}
\end{figure}

In this Section, we study the joint distribution of binary separations
and mass ratios in USco, taking into account the detection limits. The
mass  ratios are estimated  from the  magnitude differences  using the
isochrones,  as explained  above,  while the  angular separations  are
translated into linear projected separations $s = \rho d$ assuming the
common distance  $d= 140$\,pc. We consider all  binaries regardless of
their hierarchy,  i.e. include  subsystems of levels  11 and  12.  The
binaries  are   grouped  according   to  their  masses   $M_*$  and
separations  $\rho$. The {\it  companion star  fraction} (CSF)  is the
binary  fraction  in  the  corresponding group  after  accounting  for
the incompleteness.

Figure~\ref{fig:stat} plots the  separations and mass ratios overlayed
on the detection probability for the full sample.  The detection power
is sufficient to study  binaries with $q>0.3$ at projected separations
$s$ from $\sim$10 au  to 2800 au.  Figure~\ref{fig:stat} clearly shows
that  the  distribution  of  the  mass ratio  depends  on  separation.
Binaries  with $q >  0.7$ are  rare at  separations from  1\arcsec ~to
5\arcsec,  but are present  at smaller  and larger  separations.  This
effect is quantified in the following subsections.


\subsection{Distribution of the mass ratio}

\begin{figure}
\epsscale{1.1}
\plotone{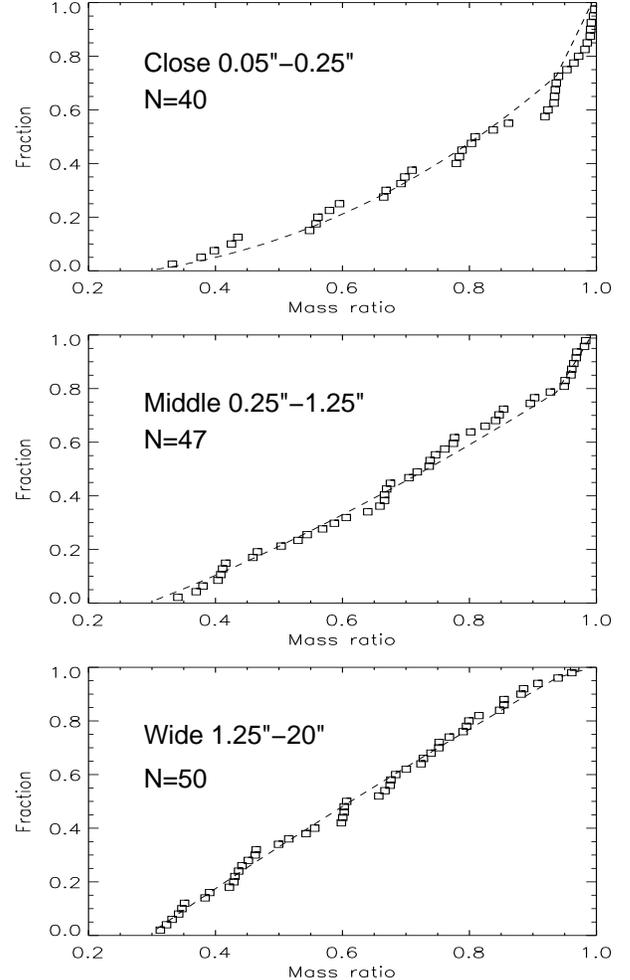}
\caption{Cumulative distributions of the mass ratio (squares) at $q>0.3$ 
and their analytical models (dashed lines) in three ranges of
separations for stars with $0.4 < M_* < 1.5$ \msun. 
\label{fig:fq}
}
\end{figure}

Following \citet{MD2017}, we model  the mass ratio distribution by the
truncated  power  law $f(q)  \propto  q^{\gamma_{0.3}}$  in the  range
$[q_{\rm  min}, q_{\rm max}]$.   Looking at  Figure~\ref{fig:stat}, we
note little  detection power for  close pairs with $q<0.3$,  and adopt
the modeled range of $[0.3, 1]$, as in the above paper.  Note that the
$q$ range  is, implicitly, also  a model parameter  and $\gamma_{0.3}$
depends on  it.  The model is  the {\em truncated} power  law, not the
general  power law.   We also  model the  separate population  of twin
binaries with  the mass ratios uniformly distributed  between 0.95 and
1.  It is characterized by the twin fraction $f_{\rm twin}$, i.e.  the
excess  of  twins with  respect  to the  power  law,  relative to  all
binaries with  $q >  0.3$, in conformity  with the definition  used by
\citet{MD2017}.  The analytic model  also includes the binary fraction
$\epsilon$ to account correctly for undetected companions.

The parameters  of the  $f(q)$ model $(\epsilon,  \gamma_{0.3}, f_{\rm twin})$
and their  confidence limits are determined by  the maximum likelihood
(ML) method,  as in \citep{FG67}.  The likelihood  function ${\cal L}$
is
\begin{equation}
{\cal L} = 2 N f_0 - 2 \sum_{i=1}^K \ln  ( f(q_i) d_i ),
\label{eq:L}
\end{equation}
where  $N$ is  the  sample size,  $q_i$  are the  mass  ratios of  $K$
binaries, and  $d_i$ are the detection probabilities  for these pairs.
The  probability of companion detection  for the complete
sample is
\begin{equation}
f_0 = \int_{q_{\rm min}}^{q_{\rm max}} f(q) d(q) {\rm d} q .
\label{eq:f0}
\end{equation}
It  is important to  realize that  the calculation  of $f_0$  uses all
stars  in  the  chosen  sub-sample,  not  only  binaries,  to  account
statistically for undetected companions. The integral in (\ref{eq:f0})
is  evaluated numerically  on  a discrete  $q$  grid.  The  confidence
limits of 68\%  (1 $\sigma$) and 90\% correspond  to the hyper-surface
of  ${\cal L}$  defined by  the increment  of 1.0  and 2.17  above its
minimum \citep{NumRec}.  The ML code was tested using simulated binary
samples filtered  by simulated incomplete detection to  verify that it
recovers known parameters of  $f(q)$. The excessive number of binaries
with $q  = 1$  resulting from the  speckle photometry bias  (excess of
$\Delta  m =  0$)  is dealt  with  by distributing  these mass  ratios
uniformly in  the interval [0.95,  1.0]. We cannot measure  these mass
ratios accurately, but know that they are close to one.

We fit the mass ratio distribution for binaries in the selected ranges
of separation and primary mass. The detection limits, computed
initially for the complete sample, are filtered accordingly to match
only the chosen sub-sample. 

\begin{deluxetable}{ l l r  r r }
\tabletypesize{\scriptsize}
\tablewidth{0pt}
\tablecaption{Parameters of the mass ratio distribution 
\label{tab:fq}}
\tablewidth{0pt}     
\tablehead{
\colhead{Mass range} &
\colhead{Sep. range} &
\colhead{$N_b$} &
\colhead{$\gamma_{0.3}$ } &
\colhead{$f_{\rm twin}$} \\
\colhead{(\msun)} &
\colhead{(\arcsec)} &
}
\startdata
0.4--1.5 & 0.05--0.25   &  40 &     1.50$\pm$0.62 & 0.15$\pm$0.08 \\
0.4--1.5 & 0.25--1.25   &  47 &     0.43$\pm$0.49 & 0.13$\pm$0.07 \\
0.4--1.5 & 1.25--20     &  50 &   $-$0.18$\pm$0.37 & $-$0.04$\pm$0.03 \\
0.4--1.5 & 0.1--1.0     &  65 &     0.60$\pm$0.42 & 0.09$\pm$0.05 \\
0.4--1.5 & 1.0--10      &  37 &   $-$0.72$\pm$0.46 & 0.00$\pm$0.04 \\
0.4--0.7 & 0.05--1.25   &  44 &    0.20$\pm$0.54 & 0.23$\pm$0.07 \\
0.7--1.5 & 0.05--1.25   &  43 &    1.50$\pm$0.55 & 0.05$\pm$0.07 \\
1.5--10  & 0.05--1.25   &  23 & $-$1.27$\pm$0.59 &    0.00$\pm$0.04 \\
0.4--1.5 & 0.1--20      & 137 &     0.41$\pm$0.28 & 0.08$\pm$0.04 
\enddata
\end{deluxetable}

The fitted  parameters of $f(q)$  are given in  Table~\ref{tab:fq} for
stars in several  ranges of separations and masses.   The third column
gives  the  number  of  pairs  $N_b$ in  the  chosen  intervals.   The
cumulative   plots  in   Figure~\ref{fig:fq}   demonstrate  that   the
analytical  models adequately describe  the data.   We note  that with
increasing  separation the  fraction of  twins decreases  and low-mass
companions  become   more  frequent,  $\gamma_{0.3}$   decreases.   We
experimented by selecting different  ranges of separation and mass and
found this  trend to be  very robust.  We  also see how $f(q)$  in the
fixed separation  range evolves  with mass: $f_{\rm  twins}$ decreases
with mass, while $\gamma_{0.3}$  increases and decreases.  When a wide
range  of separations and  masses is  selected, $f(q)$  becomes almost
uniform (see the last line in Table~\ref{tab:fq}).

\subsection{Separation distribution and companion frequency}

\begin{figure}
\epsscale{1.1}
\plotone{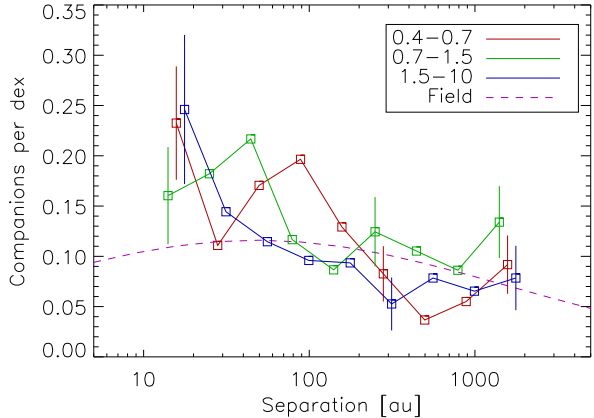}
\caption{Distribution  of  separations for  binaries  with $q>0.3$  in
  three mass intervals: 0.4--0.7, 0.7--1.5, and 1.5--10 \msun (see the
  legend box).  Each point corresponds to the  0.5-dex wide separation
  bin. The  bin limits are chosen with  a step of 0.25  dex, hence the
  adjacent points  are not independent.  The  dashed curve corresponds
  to  the field  solar-type binaries.   Representative error  bars are
  shown for three bins.
\label{fig:sephist}
}
\end{figure}

We computed the frequency of binary companions with $q>0.3$ per decade
of separation, $f_i$,  in five logarithmic separation bins  of 0.5 dex
width between 0\farcs063 and 20\arcsec ~(9 to 2800 au) by counting the
number of  companions $n_i$ and dividing  it by the  product of sample
size $N$, average detection probability $d_i$, and the bin width: $f_i
=   n_i   /  (0.5   N   d_i)$.    These   numbers  are   reported   in
Table~\ref{tab:sep}.  The  detection probability is  $\sim$0.7 only in
the first  bin; at wider  separations, the detection of  binaries with
$q>0.3$ is nearly  complete (see Figure~\ref{fig:stat}).  The fraction
of   companions    per   decade   of   separation    is   plotted   in
Figure~\ref{fig:sephist} for  three intervals of mass.   The plots are
computed  with a  sliding bin  of 0.5  dex width,  hence  the adjacent
points are not statistically independent. Typical errors are shown for
three bins.   The last  two lines of Table~\ref{tab:sep}  give the
  frequency of companions with $q > 0.3$ in the first 1-dex bin and in
  the full separation range of 2.5 dex.

For comparison, we use the log-normal separation distribution of field
solar-type binaries within  25\,pc from \citet{R10} (median $\log_{10}
(P/1d)=  5.0$,  $\sigma_{\log  P}   =  2.3$  dex,  companion  fraction
CSF=0.60$\pm$0.04), converted into  the distribution of semimajor axis
for  a system  mass  of 1.5  \msun.   The mass  ratio distribution  of
solar-type  binaries   is  almost  uniform   for  $0.05  <  q   <  1$,
independently of the period  \citep{FG67}.  Therefore, the fraction of
companions with $q > q_{\rm  lim} $ is CSF$_{\rm qlim}$=CSF$\times(1 -
q_{\rm  lim})/0.95$, or  0.44  for $q>  0.3$.   M.~Moe (2019,  private
communication)  confirmed  that CSF=0.44  is  adequate for  solar-type
binaries with  $q>0.3$.  The binary  fraction of M-type dwarfs  in the
field is taken from the  work by \citet{Winters2019}. We adopt the CSF
of  0.35$\pm$0.02 to  account for  the larger  binary fraction  in the
early-M  stars, guided  by Figure~19  of their  paper.  The log-normal
separation distribution peaking at 20\,au  with a width of 1.16 dex is
used. Considering  that M-type  binaries prefer large  $q$, we  do not
apply any  correction while comparing to M-type  binaries with $q>0.3$
in  Usco.  The  companion frequency  for early  M-type  and solar-type
field   stars   from   the   cited   publications   is   reported   in
Table~\ref{tab:sep}.

Strictly speaking, the incompleteness correction depends on $f(q)$ and
our formula is  valid only for a uniform $f(q)$;  for a rising $f(q)$,
we  over-correct.   When we  adopt  $f(q)$  with $\gamma_{0.3}=1$  and
$f_{\rm twin} = 0.15$ at  separations below 1\arcsec ~for the two first
mass bins and  repeat the calculation, the companion  frequency in the
first line  of Table~\ref{tab:sep} decreases by  $\sim$0.02, from 0.23
to 0.21 and from 0.16 to  0.14. The overall companion frequency in the
full studied separation range also  drops by 0.02.  There is no effect
at  wider separations, where  the detection  probability is  high.  We
report the results without correcting  for $f(q)$, i.e.  assume a flat
$f(q)$. 

The   projected  separation   $s$  equals   the  semimajor   axis  $a$
statistically,  the  scatter  of  $\log_{10} (s/a)$  is  $\pm$0.3  dex
\citep{FG67}, so  the distribution of projected  separations is indeed
representative of the semimajor axis distribution.  However, any sharp
features of  the latter, if they  existed, would be washed  out in the
distribution of  $s$ owing to  projections and random  orbital phases.
Conversely, any detail of the separation distribution is suspicious if
it appears only  in one 0.5-dex bin.  Considering  the projections, it
makes sense  to compute the smoothed separation  distribution by using
sliding bins, as in Figure~\ref{fig:sephist}.

\begin{deluxetable*}{c c  | rrrr |  rrrr | rrr  }
\tabletypesize{\scriptsize}
\tablewidth{0pt}
\tablecaption{Fraction of companions with $q>0.3$  per decade of separation
\label{tab:sep}}
\tablewidth{0pt}     
\tablehead{
\multicolumn{2}{c|}{Separation} &
\multicolumn{4}{c|}{0.4--0.7 \msun ~($N$=210)} & 
\multicolumn{4}{c|}{0.7--1.5 \msun ~($N$=209)} & 
\multicolumn{3}{c}{1.5--10 \msun ~($N$=153)}  \\
(arcsec) &
(au) &
$n_i$  &
$d_i$  &
\colhead{$f_i$}  &
$f_{\rm field}$  &
$n_i$  &
$d_i$  &
\colhead{$f_i$}  &
$f_{\rm field}$  &
$n_i$  &
$d_i$  &
\colhead{$f_i$}  
}
\startdata
0.06--0.20 & 16   & 17 & 0.67 & 0.23$\pm$0.06 & 0.118  & 11 & 0.66 & 0.16$\pm$0.05 & 0.106 &11 & 0.58 & 0.25$\pm$0.07  \\ 
0.20--0.63 & 50   & 18 & 0.97 & 0.17$\pm$0.04 & 0.112  & 22 & 0.97 & 0.22$\pm$0.05 & 0.114 & 8 & 0.91 & 0.12$\pm$0.04  \\
0.63--2.0  & 160  & 14 & 0.99 & 0.13$\pm$0.03 & 0.088  &  9 & 1.00 & 0.09$\pm$0.03 & 0.111 & 7 & 0.98 & 0.09$\pm$0.03  \\
2.0--6.3   & 500  & 4  & 1.00 & 0.04$\pm$0.02 & 0.058  & 11 & 1.00 & 0.11$\pm$0.03 & 0.096 & 6 & 1.00 & 0.08$\pm$0.03 \\ 
6.3--20    & 1600 & 10 & 1.00 & 0.09$\pm$0.03 & 0.032  & 14 & 1.00 & 0.13$\pm$0.04 & 0.076 & 6 & 1.00 & 0.08$\pm$0.03  \\
\hline
0.06-0.63  & 28     &  35 & \ldots & 0.20$\pm$0.04 & 0.115 &  33 & \ldots & 0.19$\pm$0.03 & 0.110  &   19 & \ldots & 0.18$\pm$0.04  \\
0.06-20    & \ldots &  63 &\ldots  & 0.33$\pm$0.04 & 0.204 &  67 & \ldots & 0.35$\pm$0.04 & 0.252  &   38 & \ldots & 0.31$\pm$0.05 
\enddata
\end{deluxetable*}

The   separation   distributions   in   Figure~\ref{fig:sephist}   and
Table~\ref{tab:sep}  show an  excess of  pairs with  separations below
100\,au relative  to the  field.  In the  9--90 au bin,  the companion
frequency  in the  field  is 0.115$\pm$0.007  and 0.110$\pm$0.007  for
early  M-type  and  solar-type  stars,  respectively.   The  companion
frequency  in  USco  in  this   bin  exceeds  that  in  the  field  by
1.75$\pm$0.35  and  1.72$\pm$0.33  times,  respectively.   The  formal
significance of both ratios,  measured independently of each other, is
$\sim$2.2 $\sigma$.  In the full  2.5 dex separation range, the excess
of  early M-type and  solar-type binaries  in USco  over the  field is
1.62$\pm$0.22 and 1.39$\pm$0.18,  respectively.  This excess therefore
seems  real,   and  with   our  significantly  larger   sample  weighs
importantly  on growing evidence  from previous  studies like  that of
\citet{Duchene2018}, based on smaller sets of stars.  It is noteworthy
that the frequency of companions does not increase with mass, contrary
to the binary statistics in the field.

\begin{figure}
\epsscale{1.1}
\plotone{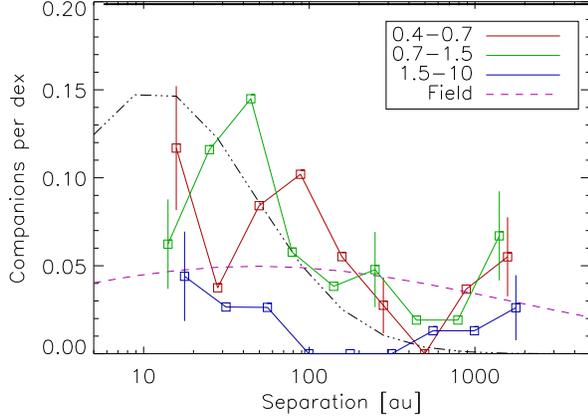}
\caption{Same as Figure~\ref{fig:sephist} for binaries with $q > 0.7$.
  The dash-dot  line is the log-normal  distribution of multi-periodic
  binaries  in   USco  found  by   \citet{TokBri2018}  with  arbitrary
  normalization.
\label{fig:sephistq}
}
\end{figure}

The dependence of the mass ratio distribution on separation means that
the  separation   distribution  also   depends  on  the   mass  ratio.
Concentration  of binaries  with  large $q$  at  small separations  is
evident  in  Figure~\ref{fig:stat}.  When  we  compute the  separation
distribution only for binaries with $q> 0.7$, the dearth of pairs with
separations   of    a   few   hundred   au    becomes   more   obvious
(Figure~\ref{fig:sephistq}).   The  excess   of  close  and  large-$q$
binaries over similar  solar-type pairs in the field  is even stronger
for  stars less  massive  than 1.5  \msun.   As for  the more  massive
binaries,  they  prefer small  $q$  and do  not  show  any excess  for
$q>0.7$.  More strikingly, we note an under-abundance of binaries with
separations of $\sim$500  au, in all three mass  ranges. However, such
binaries  with even larger  separations $s  > 1000$  au are  no longer
under-abundant.   The mass ratio  distribution at  separations between
1\arcsec ~and 10\arcsec  ~is modeled by $\gamma_{0.3} =  -0.7 \pm 0.5$
and $f_{\rm  twins}=0$ (Table~\ref{tab:fq}), indicative  of preference
for  low-$q$  binaries  and   paucity  of  large-$q$  pairs  at  these
intermediate separations. The deficit of large-$q$ binaries is further
discussed in Section~\ref{sec:disc}.

\subsection{Clustering and trapezia}
\label{sec:cluster}

\begin{deluxetable*}{l r r r c c c }
\tabletypesize{\scriptsize}
\tablewidth{0pt}
\tablecaption{Mini-clusters (trapezia)
\label{tab:trap}}
\tablewidth{0pt}     
\tablehead{
\colhead{USn}  &
\colhead{$\rho$}  &
\colhead{$\theta$}  &
\colhead{$\Delta G$}  &
\colhead{$\varpi$}  &
\colhead{$\mu^*_\alpha$}  &
\colhead{$\mu_\delta$}  \\
& 
\colhead{(arcsec)} & 
\colhead{(deg)} & 
\colhead{(mag)} & 
\colhead{(mas)} & 
\colhead{(mas yr$^{-1}$)} & 
\colhead{(mas yr$^{-1}$)} 
}
\startdata
US0123A  & \ldots & \ldots & \ldots  & 8.24$\pm$0.10 &  $-$18.5 &  $-$28.1 \\
US0123B  &  14.899& 210.3  & 5.19    & 8.02$\pm$0.11  &$-$19.7 &$-$27.4 \\
US0123Ca &  16.375& 102.3  & 5.84    & 8.02$\pm$0.09  &$-$18.2 &$-$27.0 \\
US0123Cb &  17.903& 102.2  & 7.09    & 8.20$\pm$0.14  &$-$18.5 &$-$27.9 \\
        &        &        &         &            &       &       \\
US0133A  &\ldots & \ldots & \ldots  & 6.95$\pm$0.05 &  $-$15.9 &  $-$23.6 \\
US0133B  & 8.855  & 319.2  & 3.76    & 8.99$\pm$0.22  &$-$15.6 &$-$22.2 \\
US0133C  &14.525  & 298.75 & 6.74    & 7.01$\pm$0.16  &$-$15.6 &$-$21.8 \\
        &        &        &         &            &       &       \\
US1394A & \ldots & \ldots & \ldots  & 7.22$\pm$0.10 &  $-$5.7 &  $-$27.6 \\
US1394B & 1.798  &  225.4 & 4.33    & 9.53$\pm$0.56 &  $-$7.5 &  $-$25.4 \\
US1394C & 2.888  &  359.9 & 5.92    & 6.91$\pm$0.71 &  $-$4.2 &  $-$27.5 
\enddata
\end{deluxetable*}

\begin{figure}
\epsscale{1.1}
\plotone{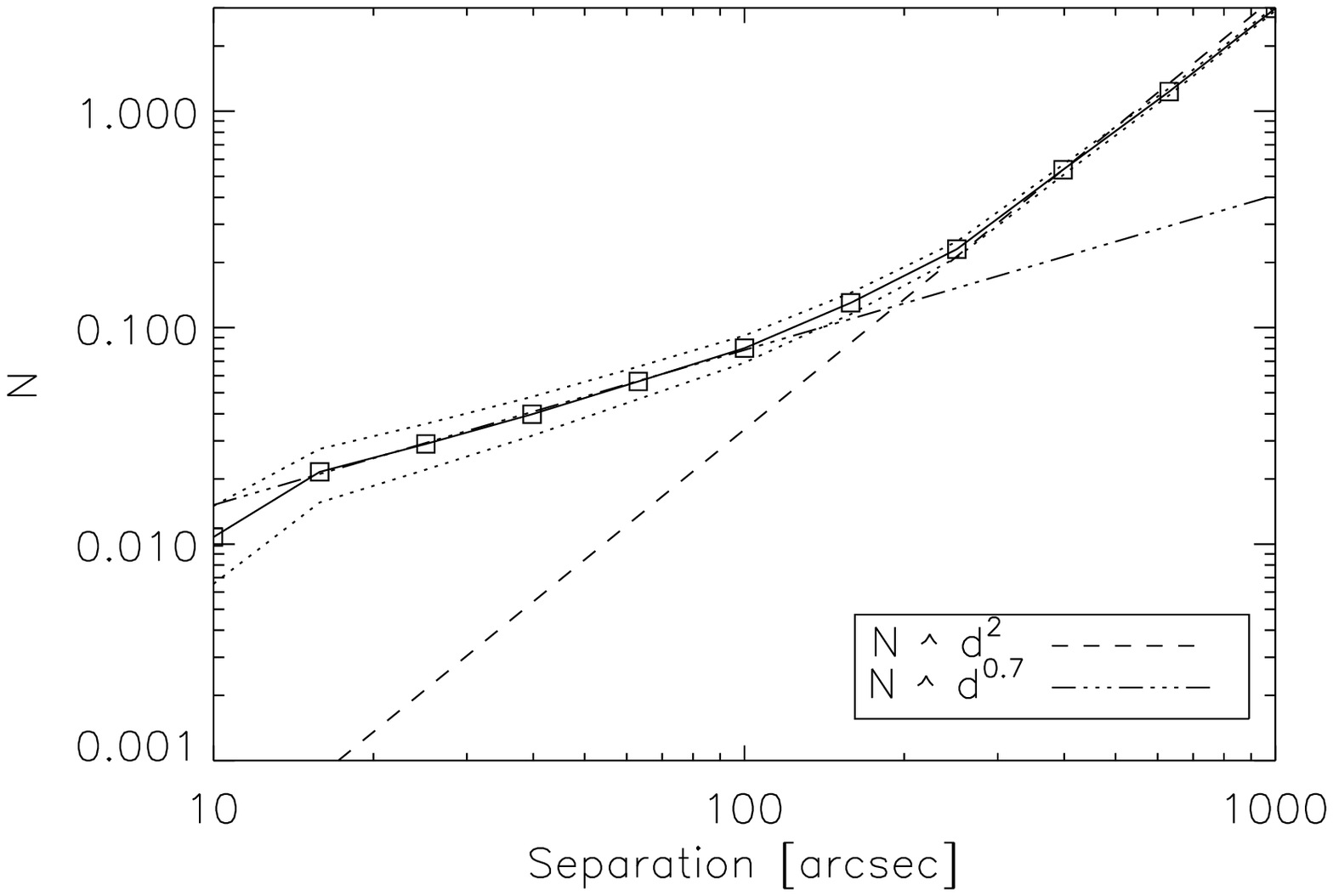}
\plotone{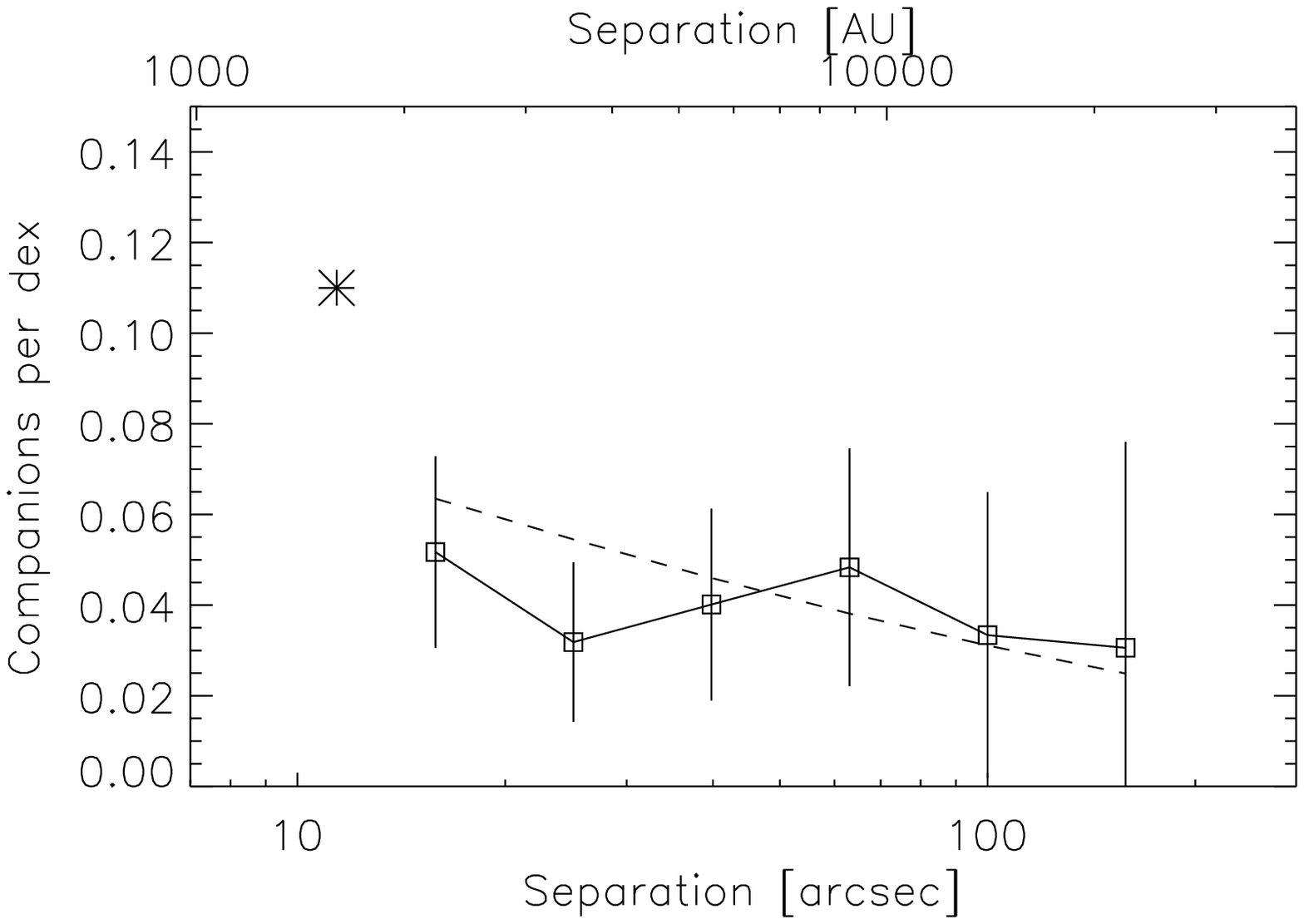}
\caption{Clustering of  the USco members.  Top: the average  number of
  companions $N$  per target within angular distance  $\rho$ (line and
  squares), with dotted lines showing the Poisson errors (they are not
  independent).  The dashed and dash-dot lines are power laws at large
  and  small  separations,   respectively.   Bottom:  distribution  of
  projected separations  after subtracting random  asterisms (line and
  squares), with Poisson error bars. The dotted line is the log-normal
  distribution in  the field  for $q>0.3$, the  large asterisk  is the
  fraction of wide companions with $q> 0.3$ from Table~\ref{tab:sep}.
\label{fig:cluster}
}
\end{figure}

To  distinguish   real  wide  binaries  from   chance  projections  of
association members,  we explored clustering of the  USco members. For
each primary  star in the filtered  sample, we computed  the number of
neighbors in the full (unfiltered)  Luhman's sample within a set of 11
angular  distances  $\rho$  from  10\arcsec ~to  1000\arcsec,  with  a
logarithmic step of  0.2 dex.  These numbers $N(\rho)$  are plotted in
the  top panel of Figure~\ref{fig:cluster}.   At $\rho>200\arcsec$,  the
growth  is  approximately  quadratic,  as  appropriate  for  a  random
distribution.  At  smaller separations, $N(\rho)  \propto \rho^{0.7}$.
The number of neighbors at close separations substantially exceeds the
extrapolated number of chance alignments.  Therefore, these pairs
are mostly  physical binaries rather than random  pairs of association
members.

In  the  bottom panel  of  Figure~\ref{fig:cluster},  the fraction  of
physical companions in each separation  bin, scaled to the bin size of
1 dex, is  plotted. The density of chance alignments  of USco stars is
estimated in  the outermost  ring, assuming that  all those  pairs are
random,  scaled  by the  relative  surface  of  the inner  rings,  and
subtracted, leaving  the number  of physical (non-random)  pairs.  The
companion frequency  is approximately  constant at $\sim$0.04  per dex
out to  $\rho \sim  80 \arcsec$ (10  kau); at larger  separations, the
increasing   statistical   errors   prevent  any   conclusions.    For
comparison,  the  dashed line  in  Figure~\ref{fig:cluster} shows  the
log-normal  distribution of  solar-type binaries  with $q>0.3$  in the
field,  and  the asterisk  is  the  frequency  of wide  $\sim$1,600\,au
binaries  with  $q>0.3$  for  stars  in  the  0.4--1.5  \msun  ~range,
estimated above.  The clustering analysis based on the Luhman's sample
does  not restrict  the range  of mass  ratios and  is subject  to the
sample incompleteness.  Therefore,  the companion frequency plotted in
Figure~\ref{fig:cluster} is   a lower  limit. All we can  say is
that  wide ($s  >10,000$ au)  binaries  in USco  and in  the field  have
similar separation distributions. 

\citet{Akter2019}  suggested  to use  the  distribution  of the  ratio
$r_{10}$ of the distances to  the nearest and 10th nearest neighbor to
distinguish  between  physical  binaries  and  chance  projections  in
samples with non-uniform  spatial distribution.  We computed distances
of 602 main targets of the  filtered sample to the members of the full
sample  and  derived  the  distribution  of  the  resulting  parameter
$r_{10}$.   It contains  the expected  signature of  physical binaries
with  a  frequency  of $\sim$0.05.  In  other words,  this  method
indicates that   $\sim$30  wide pairs  in the Luhman's  sample are
physical binaries.

Clustering in USco has been  studied previously by \citet{Kraus2008b} using
the list  of association members  available at that time.   They found
that the source density is uniform at spatial scales between 75\arcsec
~and  1\fdg7.    At  larger  scales,  not  probed   here,  the  spatial
distribution  retains  memory  of  the  primordial  clustering,  while
separations below 75\arcsec ~correspond  to binary stars.  We find the
transition from  binaries to  random projections (intersection  of the
two  lines in  Figure~\ref{fig:cluster})  at $\approx$200\arcsec  ~(30
kau).   After subtracting random  projections of  association members,
the  existence of  binaries  with separations  beyond  10\,kau is
evidenced  statistically.

A  young  stellar  cluster  is  expected to  retain  some  large-scale
structure  from  its   parental  molecular  cloud  \citep{Kraus2008b}.
Non-hierarchical   groups  of   stars  that   formed   close  together
(mini-clusters or trapezia) are dynamically unstable and disperse on a
time scale of $\sim$100 crossing times; smaller groups disperse first.
Indeed,  all three  compact non-hierarchical  asterisms found  at SOAR
(Figure~\ref{fig:trapezia})  turned out to  contain only  one physical
pair each, the other stars being unrelated objects. On the other hand,
we found  three wider non-hierarchical  groups of USco  members around
targets    US0123,   US0133,    and   US1394.     For    each   group,
Table~\ref{tab:trap} gives  relative positions, magnitude differences,
and astrometry  from GDR2.  The  parallax of US0133B differs  from the
parallaxes of the other two neighboring stars by 2\,mas, so this could
be an  association member seen  in projection onto a  14\farcs5 binary
A,C. As  for the group  US1394, the component  B is measured  at SOAR,
while the  fainter star  C is  below the SOAR  detection limit  (it is
barely detectable in  the average image).  The component  B could be a
projection, considering its slightly different parallax and PM.

The group surrounding US0123 merits special attention.  The parallaxes
and PMs  of its members are  mutually consistent.  The  component B is
identical with the target US0122, resolved at SOAR as a 0\farcs05 pair
(Figure~\ref{fig:images}).   The WDS lists  another companion  to this
star (KOH~55  at 1\farcs58)  which is  not seen at  SOAR and  in GDR2,
hence is not considered here as  real.  The components Ca and Cb, both
discovered in GDR2,  apparently form a 1\farcs5 pair  Ca,Cb. The group
thus  contains at least  five stars,  structured as  two pairs  in the
vicinity  of the  central  star of  1.7\,\msun ~mass.   Interestingly,
there are  no other USco  members within 2\arcmin, according  to GDR2,
although  a concentration  of other  targets in  this area  ($\alpha =
238\fdg82$, $\delta =-23\fdg37$) is visible in Figure~\ref{fig:sky}.

If one of the separations  in this group is shortened substantially by
projection,  this   could  be  a  hierarchical   system,  despite  its
configuration. If, on the other  hand, this is a genuine trapezium, it
could still  survive disruption, at least in  principle.  A separation
of 15\arcsec ~corresponds  to an orbital period of  $\sim$70 kyr and a
crossing time of  the same order; the age of  USco is at least 100$\times$
longer.

\section{Discussion}
\label{sec:disc}

Usually, the  distributions of binary separations and  mass ratios are
analyzed  separately \citep[e.g.][]{Duchene2013},  assuming implicitly
that  they are mutually  independent. A  more detailed  examination of
binary statistics in the field reveals that the mass ratio does depend
on the separation, and vice  versa \citep{MD2017}; close pairs tend to
have more equal components.

The  discovery of  the relative  paucity of  binaries with  $q>0.7$ at
projected separations of $s \sim 500$\,au ($\sim$ 3\farcs5 at 140 pc),
compared     to     both     smaller    and     larger     separations
(Figure~\ref{fig:sephistq}),   appears to be a  new result.  This
gap  is  present in  three  ranges  of  mass,  strengthening  its
  reality.    Binaries with smaller mass ratios  at these separations
are  not  deficient,  and   the  overall  separation  distribution  in
Figure~\ref{fig:sephist} has only a minor (if any) depression around $
s \sim 500$  au.  We verified that this effect is   unlikely to be
  produced by  a bias against semi-resolved (blended)  targets in our
input sample (see Section~\ref{sec:blend}). 

This effect explains the apparent paradox found by \citet{TokBri2018},
namely the  unusually close separations of  multi-periodic binaries in
USco.  They follow a log-normal distribution with a median of 11.6\,au
and a  dispersion of 0.6  dex.  Binaries qualify as  multi-periodic if
both components  contribute substantially to the total  light (hence a
large  $q$)  and are  an  unresolved  source  in {\em  Kepler}  ($\rho
<4$\arcsec).    Figure~\ref{fig:sephistq}   shows   that  the   narrow
log-normal separation distribution of multi-periodic binaries found by
\citet{TokBri2018}  qualitatively fits the  data at $s < 500$ au,
before the minimum.

Relative  paucity of  binaries in  USco with  separations  larger than
1\arcsec ~and equal components, compared to Taurus-Auriga, was noted by
\citet{Koehler2000}  in their Section~5.3,  but  their conclusion  was  not
statistically significant, owing to the small sample size (104 stars).
This  effect was noted  again by \citet{Laf2014}.   Meanwhile, the
compilation  of multiplicity surveys  of young  stars in  Figure~12 of
\citet{King2012} suggests  that a similar  effect may be present  in other
groups, most prominently in the Orion Nebular Cluster (ONC) studied by
\citet{Reipurth2007}, see the lower panel of Figure~\ref{fig:blend}. 

If the  narrow separation distribution found  by \citet{TokBri2018} is
extrapolated to small  separations, it would imply an  absence of very
close  (spectroscopic)  pairs.   This  is unlikely,  given  that  nine
eclipsing  binaries in  USco are  known \citep{David2019}.   No large
radial velocity  (RV) surveys of USco  have been made so  far to probe
this  regime.  \citet{Kuruwita2018} monitored  RVs of  55 disc-bearing
members of USco and  estimated the frequency of spectroscopic binaries
at $0.06^{+0.07}_{-0.02}$, like in the field and in other star-forming
regions  \citep{Duchene2013}.  Presence  of spectroscopic  binaries in
several northern  associations and  clusters (not including  USco) has
been probed by \citet{Kounkel2019}  using RV measurements from APOGEE.
They found  that the  frequency of  close binaries with  $a <  10$ au,
detectable by their survey, is compatible with solar-type stars in the
field and  does not differ substantially between  the studied regions.
Interestingly, they  discovered a deficit  of double-lined (large-$q$)
pairs   among    disc-bearing   stars,   like    those   surveyed   by
\citet{Kuruwita2018}, that is  recovered by the small-$q$ single-lined
systems with discs.  \citet{Rebull2018} also noted that multi-periodic
stars  in  USco  (large-$q$  binaries)  have a  reduced  incidence  of
discs. They  speculated that binaries destroy their  discs faster than
single  stars.    However,  the  situation  might   be  more  complex,
considering  that many  young low-$q$  spectroscopic binaries  do have
discs \citep{Kounkel2019}.

\citet{Koehler2000}  estimated the  companion star  fraction  (CSF) in
USco  in   the  separation  range   from  0\farcs13  to   6\arcsec  as
0.35$\pm$0.06, a  factor of 1.6$\pm$0.3 larger compared  to the field.
Most notably, their  Figure~7 shows that the CSF  {\em increases} with
decreasing mass, contrary to the trend in the field.  We confirm these
findings, especially for binaries with $q>0.7$.  We find that the mild
excess of binaries in USco compared to the field,  1.39$\pm$0.18 for
  solar-type  stars  and  1.62$\pm$0.22  for  early  M-type  stars,  is
produced  by  pairs  with $s<  100$  au  and  a  large $q$;  at  wider
separations, the binary frequency is similar or even lower than in the
field.

\citet{Kraus2008a}  studied multiplicity in  USco for  a sample  of 82
stars  with  masses  from 0.3  to  1.7  \msun  ~observed at  the  Keck
telescope with a  high angular resolution and a  high contrast, in the
$K$ band.  Using also published data and seeing limited imaging for 51
pairs, they derived a  CSF of $0.35^{+0.05}_{-0.04}$ in the separation
range of  6--435 au (1.9 dex),  independently of the  primary mass and
only marginally  (1.3 times)  larger than the CSF  of solar-type
stars in  the field, 0.27 (for  all $q$).  However, the  small size of
their sample precluded a  more detailed characterization of the binary
statistics.   \citet{Kraus2008a} also  noted the  paucity  of binaries
with masses in the 0.25--0.7  \msun ~range at $s>200$ au.  Figure~6 of
\citet{Duchene2018} shows  that the excess of young  binaries over the
field is most  prominent at separations $s < 100$  au; they found such
an excess  in the Orion Nebular  Cluster (ONC) from  eight close pairs
discovered by their survey.

Unfortunately, many multiplicity surveys of young stars do not specify
detection limits in terms of mass ratio or do not estimate $q$ at all,
giving  as an excuse  uncertain isochrones  and ages,  variability, or
infra-red  excess.   The  derived  companion fractions  are  therefore
difficult to  compare between themselves or with  the field. Moreover,
previous multiplicity  surveys (including those in  USco) covered wide
ranges of stellar masses because modest samples did not allow to probe
the dependence of multiplicity on mass.

Solar-type binaries in  the field have a uniform  distribution of mass
ratios that  does not depend on the  period \citep{R10,FG67}. Although
\citet{MD2017} found $\gamma_{0.3} \approx -0.5$ for the 25-pc sample,
the best-fitting  single power law in  the full range  $0 < q <  1$ is
$\gamma \approx 0$.   In contrast, mass ratios of  binaries in USco do
depend on their separation, for all masses.  Pairs with $s \sim 10-20$
au have  $\gamma_{0.3} \approx 1.5$  and a non-negligible  fraction of
twins;  $f(q)$  becomes nearly  uniform  only  at larger  separations.
Also,  the deficit  of binaries  with $q>0.7$  and  separations around
500\,au  noted in USco is  not observed in the field. Finally, in
USco the  companion fraction does  not decrease with  decreasing mass.
These differences indicate that binary statistics is not universal and
that  star-formation regions produce  binary populations  with varying
properties.   The same conclusion  was reached  by \citet{Duchene2018}
for the ONC and by \citet{King2012} who compared multiplicity in seven
young groups, including USco. They noted an excess of close ($s < 100$
au) binaries in  all young populations.  These hard  binaries  are
  unlikely to  be destroyed   by dynamical  interactions.  Therefore,
the multiple star formation (and,  by extension, the star formation in
general) is not universal.

\section{Summary}
\label{sec:sum}

The main results of our survey are:
\begin{itemize}
\item
The sample  of 614 members of  USco more massive than  0.4 \msun ~from
the list of \citet{L18} has been observed with a spatial resolution of
0\farcs05, expanding  by an  order of magnitude  previous multiplicity
surveys of this  region. We discovered 55 new pairs.  Some of them are
good candidates for future determination of orbits and masses.

\item
New observations, published surveys, and {\em Gaia} DR2 astrometry are
combined to produce the catalog  of 250 physical binaries in USco with
separations  up to  20\arcsec.  Limits  of companion  detection around
each target are quantified.

\item
We found that the mass  ratio distribution $f(q)$ depends on
the separation.  The distribution at $q>0.3$ is modeled by a truncated
power  law with  index $\gamma_{0.3}$  and an  additional  fraction of
twins with $q>0.95$.  For stars with masses between 0.4 and 1.5 \msun,
the  power index  changes from  $\gamma_{0.3} =  1.5 \pm  0.6$  in the
projected separation range from 7 at  70 au to $\gamma_{0.3} = 0.4 \pm
0.5$ in the  intermediate (70--175 au) range, to  $\gamma_{0.3} = -0.2
\pm 0.4$ in  the (175--2800 au) range. At the  same time, the fraction
of twins decreases from 0.15$\pm$0.08 to zero.

\item
The distribution of separations and the companion fraction are broadly
compatible with  those of solar-type stars  in the field,  with a mild
excess  of pairs at  separations $<$100\,au.   However, unlike  in the
field, there is no dependence of  the CSF on the stellar mass.  In the
separation range  from 9 to  2800 au we measure  CSF=0.35$\pm$0.04 for
$q>0.3$ and  masses between 0.7  and 1.5 \msun  ~and CSF=0.33$\pm$0.04
for masses from 0.4 to 0.7 \msun.  For comparison,  solar-type and
  early  M-type stars  in  the field  have  a CSF  of  0.110 and  0.115,
  respectively, in the same range of separations and $q$. In both mass
  ranges, the excess over the  field is established with a statistical
  significance exceeding 2$\sigma$.  

\item
We  discovered a  deficit  of binaries  with  $q>0.7$ at  intermediate
separations  from 200 to  500 au;  such binaries  are present  at both
smaller and larger  separations. The deficit is seen  for stars of all
masses.  It explains the unusually compact distribution of separations
found by  \citet{TokBri2018} for multi-periodic  stars; those binaries
with large $q$ are mostly closer than 1\arcsec ~owing to the deficit at
larger separations.   This effect, not present in  the distribution of
the  field binaries,  might be  discernible in  other groups  of young
stars.

\item
The multiplicity statistics in USco differs from the field in several
important aspects. 


\end{itemize}

\acknowledgements

Our survey  would have been  impossible without the  technical support
provided  by the SOAR  telescope team  and by  the CTIO  engineers; we
thank all  those involved. We  also appreciate the allocation  of some
engineering time for  this survey by the SOAR  director, J.~Elias. The
detector  of  HRcam  was kindly  loaned  by  N.  ~Law.   We  thank
the  anonymous  Referee for helping  us to  improve the  presentation and
  sharpen our conclusions. 


This  work used  bibliographic references  from the  Astrophysics Data
System maintained  by SAO/NASA and the Washington  Double Star Catalog
maintained at USNO.   We relied heavily on the  data from the European
Space        Agency       (ESA)        mission        {\it       Gaia}
(\url{https://www.cosmos.esa.int/gaia}  processed  by  the  Gaia  Data
Processing      and     Analysis      Consortium      (DPAC,     {\url
  https://www.cosmos.esa.int/web/gaia/dpac/consortium}).   Funding for
the DPAC has been provided by national institutions, in particular the
institutions participating in the Gaia Multilateral Agreement.

{\it Facilities:} \facility{SOAR}

\appendix

\section{Comments on individual binaries}
\label{sec:cmt}

Some pairs  listed in our binary  catalog (Table~\ref{tab:comp}) merit
individual  comments. The comments  are assembled  below in  a tabular
format. Each  binary is identified by  its WDS-style code  and the USn
number.


\begin{deluxetable}{ l r l }
\tabletypesize{\scriptsize}
\tablewidth{0pt}
\tablecaption{Notes on individual binaries
\label{tab:notes}}
\tablewidth{0pt}     
\tablehead{
\colhead{WDS} &
\colhead{USn} &
\colhead{Text of the note} 
}
\startdata
15322-2158 &   0 & KOU~39 (0\farcs69, $\Delta K=3.8$  mag) is not confirmed at SOAR, below detection limit of KOU05, ignored. \\
15360-2324 &   5 & KSA~114AB (0\farcs055, $\Delta K=3.0$ mag) is not resolved at SOAR, below our detection limit. \\
15415-2521 &  15 & LAF~8 at 3\farcs7 is not seen at SOAR and by {\it Gaia}, ignored. \\
15481-2513 &  44 & HDS~2226 has an orbit with a period of 31.2\,yr. \\
15522-2141 &  84 & The {\it Gaia} companion at 4\arcsec is outside the SOAR field. A faint companion at 2\farcs6, 110\degr is assumed  optical. \\
15527-2705 &  89 & A triple system: A,B at 18\farcs2 ({\it Gaia}) and Aa,Ab at 0\farcs66 (SOAR). \\
15536-2520 &  97 & SOAR does not confirm the 0\farcs1 occultation pair, only the 2\arcsec ~binary BU~36AB is seen. \\
15539-2432 & 100 & OCC~161 at 0\farcs1, discovered in 1932, is not confirmed at SOAR. \\
15553-2322 & 122 & KOH~55 at 1\farcs5 is not confirmed, but a new 0\farcs05 pair is discovered. This is the secondary \\
           &     &  component to US0123, at 14\farcs9 and 5.2 mag brighter. \\
15558-2512 & 133 & Trapezium  with companions at 8\farcs9 and 14\farcs9. The 8\farcs9 pair with a slightly different parallax is ignored. \\
15580-3144 & 193 & HNK~4 (0\farcs1,  $\Delta K=3.0$ mag) is tentatively  resolved at SOAR well below the detection limit. \\
15592-2606 & 239 & KSA~78AB is triple because the secondary, at 2\farcs9, is resolved as a 0\farcs07 pair B,C. \\
16000-2221 & 264 & A new SOAR triple (A,B at 0\farcs30 and B,C at 0\farcs07). The 0\farcs025 pair KSA~122 is ignored, too close.  \\
16001-2027 & 288 & KOH~63 has a retrograde motion of 10\degr in one year, candidate for an orbit. \\
16022-2241 & 349 & The pair LAF~94AC (0\farcs33, $\Delta K=4.0$ mag) is unresolved at SOAR, below the contrast limit. \\
16030-2257 & 378 & KSA~81 (1\farcs2,  $\Delta J=2.7$ mag) is unresolved at SOAR (contrast $\Delta I>3.0$ mag) and by {\it Gaia}. \\
           &     & The discovery measure appears to be below the claimed detection limit, hence we ignore this pair. \\
16034-1752 & 390 & A new triple: Aa,Ab at 0\farcs1 (SOAR) and A,B = KSA~82 at 2\farcs5, measured by {\it Gaia} and SOAR. \\
16039-2032 & 413 &  KOH~70 has an orbit with $P=52$\,yr \citep{TokBri2018}. \\
16040-1751 & 417 & The 2\farcs2 pair with $\Delta G = 4.3$ mag is measured by both {\it Gaia} and SOAR.  \\
           &     &  Negative GDR2 parallax for B. The pair moved by 76\,mas and is considered here as optical. \\
16044-2131 &  432 & A star at 16\farcs2, 1.7 mag brighter, USco member, is found in GDR2, but not in the sample. Ignored. \\
16048-1749 & 447 & Trapezium. The 2\farcs85 companion is physical, another at 3\farcs5 is optical, as well as two other fainter stars. \\
16048-1930 & 446  & MET~69Aa,Ab at 0\farcs04 has two resolutions in the literature, unresolved at SOAR (too close?). \\
16054-1948 & 482 & A massive quadruple; the orbits of BU~947AB and MCA~42CE (US0483), at 13\farcs7,  are known. \\
16057-2150 & 501 & LAF~104AC at 0\farcs13,  $\Delta K=2.0$ mag, is not confirmed by SOAR, closed down? Accepted as real. \\
16061-2336 & 521 & RAS~25 (0\farcs1, $\Delta I=3.3$ mag) is below SOAR detection limit. Owing to {\it Hipparcos} acceleration, considered  real. \\
16070-2033 &  569 &  KSA~85BA (11\farcs8): B is a 0\farcs066 pair TOK~744CD, brighter than A = US0571. \\ 
16075-2546 &  613 & A {\it Gaia} companion at 17\arcsec, 0.52 mag brighter, not in the Luhman's sample, ignored. \\
16082-1909 &  651 & KSA~127AB (0\farcs025) is below the SOAR limit, unresolved. \\
16084-1930 &  660 & A {\it Gaia} pair at 13\farcs4, the secondary is US0663. \\
16087-2341 & 686 & OCC~103 (0\farcs1 in 1930) is not confirmed at SOAR, considered spurious. \\
16087-2523 & 685 & A new 0\farcs05 subsystem Aa,Ab in JNN~221 is detected in 2018.25, but unresolved (or marginally) in 2019. \\
16090-1900 & 708 & HNK~6Aa,Ab (0\farcs18) is not confirmed by KOH2000, LAF14, and at SOAR,  ignored here. \\
           &     & In contrast, the wider 0\farcs96 pair KOU~55AB is resolved in 2018.56 and unresolved in 2019.2, variable? \\
16093-1927 & 728 & KOU~55 (0\farcs88) has a new subsystem Aa,Ab at 0\farcs09, $\Delta I=3.6$ mag, below SOAR detection limit? \\
16104-1904 & 797 & KOH~76AB (4\farcs6, 6\fdg5) is confirmed by GDR2 as physical, but the 4\farcs5 pair KSA~133AE is \\
           &     & optical because it moves too fast. GDR2 gives no parallax for E. \\
16104-2306 & 801 & KOU~56AB with $\Delta K = 3.2$ mag is unresolved at SOAR but confirmed by LAF14. \\
           &     & Another faint star LAF~123AC at 3\farcs06 is ignored, likely optical. \\ 
16105-1913 & 816 & KSA~93Aa,Ab  with $\Delta K = 3.0$ mag is unresolved at SOAR but confirmed by LAF14. \\
           &     & The companion B at 5\farcs8 is physical, this is a triple system. \\
16107-1917 & 820 & Discordant photometry: $\Delta G =3.19$ mag, $\Delta I= 4.0$ mag, variable? \\
16120-1907 & 907 & KOH~78AB has a variable companion? $\Delta K \sim 0$ in 1999, $\Delta R = 3.4$ in 2015, $\Delta G = 3.5$ mag in 2015.5. \\
16120-1928 & 908 & BU~120AB and CHR~146Aa,Ab is a known triple system. \\
16125-2332 & 934 & A new triple. The companion at 1\farcs6 is assumed physical, it is too faint for {\it Gaia}. \\
16127-1859 & 945 & GHE~20Aa,Ab (0\farcs12) is securely unresolved at SOAR, despite $\Delta K =1.5$ mag and several measures.  \\
	   &     & This is companion to US0947, at 19\arcsec.  Considered as independent in the statistics.  \\
16128-1801 & 957 & The 3\farcs2 Gaia pair is below the SOAR detection limit (too wide and too faint). \\
16130-2245 & 964 & Trapezium asterism. The 5\farcs4 companion is physical, the 3\farcs6 one is optical. \\
16133-2922 & 985 & Close on the sky is KOH~70 (EPIC 20252205) with a $P=52$ yr orbit, missed in the Luhhman's sample. \\
16140-2815 & 1019 & A classical binary RST~1883, orbit calculation is possible. \\ 
16156-2622 & 1107 & {\it Gaia} measured $\Delta G = -0.05$ mag.  Also measured at SOAR. Set  $\Delta G = 0.05$ mag. \\
16160-2325 & 1123 & The 2\farcs5 SOAR pair with $\Delta I =3.3$ mag is not present in {\it Gaia} and ignored. \\ 
16164-2459 & 1146 & KOU~59 is unresolved at SOAR, 0\farcs94,  $\Delta K =4.4$ mag,  below the detection limit. \\
16188-2328 & 1248 & The {\it Gaia} companion at 8\farcs5,  $\Delta G = -0.23$ mag, is accepted with  $\Delta G = 0.23$. \\
16193-2329 & 1273 & Triple asterism in SOAR and {\it Gaia}.  The 0\farcs9 pair A,B is physical, the 1\farcs9 one is optical. \\
16205-2007 & 1327 & B~1808AB, discovered at 0\farcs2  in 1929, is now at 0\farcs07, opening. The outer pair SHJ~226AC is at 12\farcs6.  \\
16209-2254 & 1345 & The secondary at 9\farcs9 is US1344. \\
16212-2536 & 1351 & The 3.1-mas interferometric subsystem NOR~1Aa,Ab is ignored, outside the surveyed range. \\
16212-2342 & 1353 & SOAR and {\it Gaia}  measure the 1\farcs6 pair at different positions, discordant parallax, optical. \\
16222-1953 & 1394 & {\it Gaia} trapezium: A,B at 1\farcs8, A,C at 2\farcs89.  A,B is confirmed at SOAR, A,C is barely seen.  \\
16239-3312 & 1443 & The companion at 6\farcs2 is US1444. \\
16263-2233 & 1486 & The secondary at 10\farcs5 is US1487. \\
16298-2152 & 1531 & MET~77AB (3\farcs1, $\Delta K =5.8$ mag) is fixed in 2002-2012, physical. Not seen by {\it Gaia}, barely by SOAR.  \\
16320-2530 & 1551 & The secondary at 14\farcs2 (BOV~58) is US1553. \\
16336-1833 & 1562 & The faint {\it Gaia}  companion at 3\farcs3 is at the edge of the SOAR image. \\
16359-2813 & 1578 & The 21-mas pair RIZ~18 is below the SOAR resolution limit. 
\enddata
\end{deluxetable}

\end{document}